\documentclass[12pt]{article}

\date{}

\usepackage{amsmath}
\usepackage{graphicx}

\title{ Di-Pion  Decays of Heavy Quarkonium in the Field Correlator Method.}

\author{Yu.A.Simonov\\
 State Research
Center\\Institute of Theoretical and Experimental Physics, \\
Moscow, 117218 Russia}
\newcommand{\beq}{\begin{eqnarray}}
 \newcommand{\eeq}{\end{eqnarray}}
\newcommand{\be}{\begin{equation}}
 \newcommand{\ee}{\end{equation}}

\newcommand{\begat}{\begin{gathered}}
 \newcommand{\eegat}{\end{gathered}}

\def\fun#1#2{\lower3.6pt\vbox{\baselineskip0pt\lineskip.9pt
\ialign{$\mathsurround=0pt#1\hfil ##\hfil$\crcr#2\crcr\sim\crcr}}}
\newcommand{\veX}{\mbox{\boldmath${\rm X}$}}
\newcommand{{\SD}}{\rm SD}

\newcommand{\vex}{\mbox{\boldmath${\rm x}$}}
\newcommand{\vey}{\mbox{\boldmath${\rm y}$}}
\newcommand{\ver}{\mbox{\boldmath${\rm r}$}}

\newcommand{\veP}{\mbox{\boldmath${\rm P}$}}
\newcommand{\vep}{\mbox{\boldmath${\rm p}$}}
\newcommand{\veK}{\mbox{\boldmath${\rm K}$}}
\newcommand{\veq}{\mbox{\boldmath${\rm q}$}}

\newcommand{\vez}{\mbox{\boldmath${\rm z}$}}

\newcommand{\veL}{\mbox{\boldmath${\rm L}$}}
\newcommand{\veR}{\mbox{\boldmath${\rm R}$}}

\newcommand{\vek}{\mbox{\boldmath${\rm k}$}}

\newcommand{\veu}{\mbox{\boldmath${\rm u}$}}
\newcommand{\vev}{\mbox{\boldmath${\rm v}$}}

\newcommand{\verho}{\mbox{\boldmath${\rm \rho}$}}

\newcommand{\vepi}{\mbox{\boldmath${\rm \pi}$}}
\newcommand{\veta}{\mbox{\boldmath${\rm \eta}$}}

\newcommand{\veE}{\mbox{\boldmath${\rm E}$}}

\newcommand{\lan}{\langle}
\newcommand{\ran}{\rangle}
\begin{document}
\maketitle

\begin{abstract}

Mechanism of di-pion transitions $nS\to n'S\pi\pi(n=3,2; n'=2,1)$
in bottomonium and charmonium is studied with the use of the
chiral string-breaking Lagrangian allowing for the emission of any
number of $\pi(K,\eta),$ and  not containing fitting parameters.
The transition amplitude contains two terms, $M=a-b$, where first
term (a) refers to subsequent one-pion emission:
$\Upsilon(nS)\to\pi B\bar B^*\to\pi\Upsilon(n'S)\pi$  and second
term (b) refers to two-pion emission: $\Upsilon(nS)\to\pi\pi B\bar
B\to\pi\pi\Upsilon(n'S)$. The one-parameter formula for the
di-pion mass distribution is derived, $\frac{dw}{dq}\sim$(phase
space) $|\eta-x|^2$, where
$x=\frac{q^2-4m^2_\pi}{q^2_{max}-4m^2_\pi},$ $q^2\equiv
M^2_{\pi\pi}$.
 The parameter  $\eta$ dependent on the  process is calculated, using SHO
 wave functions and imposing PCAC restrictions (Adler zero) on
 amplitudes a,b. The resulting di-pion mass distributions  are in agreement with experimental data.

\end{abstract}

\section{Introduction}


An enormous amount of experimental data on strong decays of mesons
and baryons is partly used by theoreticians for comparison in the
framework of the $~^3 P_0$ model \cite{1}, and its flux-tube
modifications \cite{2}. The analysis done in \cite{3} confirmed
the general validity of the model, whereas in \cite{4} results of
other forms of decay operators have been also investigated in
meson decays, and in \cite{5} in baryon decays. On the whole, the
phenomenological picture seems to be satisfactory for the $~^3
P_0$ model with some exclusions discussed in \cite{3} and
\cite{6}. The  extensive study of strong decays of strange
quarkonia  based on the $~^3P_0$ model  was done in \cite{7}, for
 a history and review of strong decays see \cite{8,9}.

 Another form of channel coupling Lagrangian was used by the
 Cornell group \cite{10}, and recently exploited in higher
 charmonia levels \cite{11}.

Of special interest are the OZI allowed strong decays of heavy
quarkonia, e.g. 1) $\psi(n)\to D\bar D,$ $\psi(n) \to D^*  \bar
D^*$ etc. and 2) $\psi(n)\to D\bar D(\pi\pi),$  or  3) $\psi(n)
\to D \bar D \pi(\eta)$. Pionic decays  are especially important
since they provide a fundamental  probe of decay mechanism.

The dipionic decays  of heavy quarkonia were first discovered in
\cite{12},  and  since  then attracted a lot of attention, because
 they display characteristic process
of a pion pair creation from the nonperturbative gluonic vacuum.

The resulting dipion spectrum was reconstructed  from the general
requirements of Lorentz invariance and PCAC \cite{13,14} and the
multipole expansion of gluonic field was used as derived in
\cite{14, 15,16,17}.

In this derivations it was assumed that decaying heavy quarkonium
has a small radius  so that the multipole expansion is applicable.
In practice, however, the radii of $\Upsilon(2S), \Upsilon(3S),
\Upsilon(4S), \Upsilon(5S)$, are respectively 0.5,0.7,0.9, 1.1 fm
while these of $\psi (2S)$ and $\psi (3770)$ are 0.8 fm,
\cite{18}. These values are larger than the gluonic correlation
length of the vacuum, $\lambda= 0.2$ fm \cite{19,20}, therefore
not the field strength but rather string tension comes into play
and the internal structure of the decaying state can be important.
Therefore one should use another and  more complete formalism to
calculate the transitions. As a possible hint in this direction
may serve the anomaly in the pionic spectrum. Indeed, while  the
 $\Upsilon(nS)\to \Upsilon(n'S)\pi\pi$ transitions with $n=2, n'=1$, observed by CLEO
\cite{21,22,23,24}, show a high mass peak in $\pi\pi$ spectrum,
the $\Upsilon(3S)\to\Upsilon(1S)\pi\pi$ transition exhibits a
double peak structure. The same type of structure was found by
BaBar in
 $\Upsilon(4S) \to \Upsilon(2S)\pi\pi$ \cite{25}, while both Belle \cite{26}
 and BaBar see only high mass peak in $\Upsilon(4S) \to \Upsilon(1S)\pi\pi$.
Some  theoretical explanations of this anomaly were suggested in
literature a) the role of final state interaction and $\sigma$
resonance \cite{27,28,29,30}, exotic $\Upsilon-\pi$ resonances
\cite{27,31,32,33}, coupled channel effects \cite{34,35},
relativistic corrections \cite{36}, S-D mixing \cite{37}. The role
of the constant term was studied in \cite{38}.

Another interesting example of the important role of $(Q\bar Q)
\pi\pi$ channel is provided by the $X(3872)$ resonance, which was
seen in this channel \cite{39} as well  as in the channel $(D\bar
D\pi)$ \cite{40}, and by recent finding of  the  $Y (4260)$
resonance in the channel $J/\psi \pi^+\pi^-$\cite{41}.

Recently a detailed analysis of dipion transitions among
$\Upsilon(3S),$ $\Upsilon(2S),$ $\Upsilon(1S)$ states was done by
CLEO Collaboration  \cite{42}, which gives additional information
on mass and $\cos \theta$ distributions, calling  for theoretical
explanation.

From all this set of accumulated data one derives the  impression
that one should construct the theory, where large distances are
under control and all channels like $D\bar D, D\bar D^*, D\bar D
\pi,  D\bar D\pi\pi$ for $c\bar c$ and similar ones  for $b\bar b$
states should enter on the same ground. Moreover one  should try
to derive it from QCD with minimal number of assumptions and
parameters. As a result the theory must be applicable to large
distances and radii of bound states,  $R> \lambda =0.2$ fm.

It is the purpose of the present paper to make some progress in
this direction using the Field Correlator Method (FCM) \cite{43}
and background perturbation theory \cite{44} to treat
nonperturbative (NP) QCD contributions together with perturbative
ones.

Physically, the  main mechanism is the breaking of string,
connecting heavy $Q$ and $\bar Q$,  by a light $q\bar q$ pair with
a possible emission of Nambu-Goldstone mesons. As we shall argue
below, there exists a general mechanism for the processes of this
kind, which is derivable from QCD using FCM, along the lines first
treated in \cite{45}.

To this end we shall find the Green's function of heavy $Q$ and
$\bar Q$ quarks, propagating in the nonperturbative gluonic
vacuum, where creation of the light $q \bar q$ pair is described
by the quark-pion Lagrangian (action) $S_{QM}$ obtained via the
chiral bosonization procedure \cite{46}. It is essential, that our
derivation does not contain any arbitrary parameters, and the only
mass parameter entering $S_{QM}, M_{br}$  can be in principle
computed via field correlators. The resulting structure of the
decay amplitude for the process  (1) resembles to some extent the
$~^3P_0$ model, and this fact can be used as a step  in the
systematic construction of the theory of strong decays, where
$~^3P_0$ (or its modifications) is participating as an  example.

The paper is organized  as follows. In  section 2 the general
formalism of field correlators for the  $Q \bar Q$ Green's
function and for the effective Lagrangian of light quarks is
introduced. In section 3 the bosonization procedure for the
effective Lagrangian is described and the  effective quark-pion
operator is written down. In section 4 general relativistic
construction of decay amplitude is given. Section 5 is devoted to
the quarkonia decays without pions. In section 6 the dipion decays
are considered and the expressions of decay amplitude, dipion
spectra and total width are given. Section 7 is devoted to results
of analytic and numerical calculations and comparison to
experiment. Section 8 contains discussion of results in comparison
to other approaches and experiment. Section 9 is devoted to
summary of results and outlook. Acknowledgements are placed in
section 10. Four appendices cover necessary technicalities for
relativistic decay amplitude, kinetics and details of SHO matrix
elements.

\section{Effective quark-pion Lagrangian}

We start here with the standard Euclidean partition function for 3
light flavors of  quarks with mass matrix $\hat{m} = \left(
\begin{array}{lll} m_1&&\\&m_2&\\&&m_3\end{array}\right),$ and $\psi$ operators
$\psi^g, g=1,2,3$  and one heavy flavor with mass $m_Q$ and wave
operator $\Psi_Q$ \be Z= \int DA D\psi^g D\bar\psi^g D\Psi_Q D\bar
\Psi_Q \exp [-(S_0 + S_1+ S_{int} + S_Q)].\label{s1}\ee

We have omitted in (\ref{s1}) gauge-fixing and ghost terms, and
defined \be S_0 =\frac14 \int d^4 x  (Fâ_{\mu\nu})^2, ~~ S_1 =-i
\int d^4 x \bar \psi^f (\hat \partial + m_f) \psi^f,\label{s2}\ee
$$S_{int}=-\int d^4 x \bar \psi^f g\hat{A}\psi^f,~~ S_Q =-i\int d^4 x\bar \Psi_Q (\hat D + M_Q) \Psi_Q.$$

To derive the quark-meson Lagrangian one can follow the procedure
given in \cite{46}. The first step is the integration  over
gluonic fields $DA$ using the generalized contour gauge  \cite{47}
to express $A_\mu$ through $F_{\mu\nu}$ and the Gaussian
approximation to cluster expansion as in \cite{43}.

In this way one has \be A_\mu (x) = \int_{L(x,x_0)} d
\Gamma_{\mu\nu\rho} (x,z) F_{\nu\rho} (z) \label{s3}\ee where the
contour $L(x, x_0)$ connects some arbitrary initial point $x_0$
and final point $x$ and the measure $d\Gamma_{\mu\nu\rho}$ is
given in \cite{46} and will not be used here.

The partition function $Z$ does not depend on the choice of
contour $L$, and one can introduce the unity operation of
integration over some class of contours $\{L\}$ with the weight
$D\kappa(L)$ In the Gaussian approximation in the confining regime
the area $S_{Q\bar Q}$ inside closed world lines of quarks and
aintiquarks will depend on the choice of contours $\{C\}$, and we
shall assume that the contour integration $D\kappa(L)$ results in
the fixing of the minimal surface $S^{min}_{Q\bar Q}$. Actually
this minimal surface appears in the area-law asymptotics of Wilson
loop when all field correlators (and not only Gaussian ones) are
taken into account, and hence integration $D\kappa(L)$ restores
the action of all correlators.

In what follows we shall be interested in the processes of  light
quark creation, where heavy quarkonium is involved as a
background. To this end it is  convenient to integrate
$\Phi_f\Phi_{in}$ $D\Psi_Q D\bar \Psi_Q$ with $\Phi_{in,f} =
(\Psi_Q\Gamma_{in,f} \Psi_Q)$ before integrating over $DA$, which
yields \be Z=\int DA D\psi D\bar \psi e^{-(S_0+S_1+S_{int})}G_{Q
Q} (x_{in}, x_f; A) \label{s4}\ee
 where we have defined
 \be G_{QQ}
(x_{in}, x_f; A)= G_{QQ}^{(conn)} + G_{QQ} ^{(disc)} \label{s5}\ee
with
 \be G_{QQ}^{(conn)}= \frac{1}{N_c} tr \{\Gamma_f S_Q (x_{in},
x_f;A) \Gamma_{in} S_Q (x_f, x_{in}; A)\}. \label{s6}\ee
 Here
$G_{QQ}^{(disc)}$ refers to the OZI violating decay and production
processes, which will  not be discussed in the present paper, and
this term will be omitted in what follows.

The connected part of $G_{QQ}$ can be written using the
Fock-Feynmann-Schwinger (FFS) path-integral representation
\cite{48} \be G_{QQ}^{(conn)} (x_f, x_{in}; A) =\int d\rho (C_Q)
W(C_Q; A) \label{s7}\ee where the Wilson loop operator  is defined
as
 \be
W(C_Q,A) =\frac{1}{N_c} tr_c P_A(\exp (ig \int_{C_Q} A_\mu
dz_\mu)) \label{s8}\ee and the $d\rho(C_Q)$ includes integration
over paths $C_1, C_2$ of  heavy quark and antiquark $d\rho (C_Q)
\sim Dz D\bar z$, so that $C_Q = C_1+ C_2.$

The next step is the integration over all gluonic fields in
(\ref{s4}) which reduces to the expression $\int DAe^{-S_{int} }
W(C_Q, A)$. Here $A_\mu$ enters linearly in the exponent and one
can use the cluster expansion theorem and keep only the quadratic
terms in $A_\mu$ (and due to (\ref{s3}) also quadratic in
$F_{\mu\nu}$) -- this approximation which is usually called the
Gaussian one -- is justified by the Casimir scaling accurate to
few percent  \cite{49}.

The result of this integration can be written  as (see \cite{50}
for details of derivation)
$$\int DAe^{(S_{int}(A) +S_0 (A))} W(C_Q,
A) \equiv \lan e^{-S_{int} (A)} W(C_Q, A)\ran$$\be= \int D\kappa
(L) e^{-S_{eff} (L) }\bar W (C_Q,L). \label{s9}\ee
 Here $S_{eff}(L)$
is the same as in the case of light quarks without heavy quarkonia
 and $\bar W
(C_Q, L)$ are expressed through the same field correlator.
 \be
S_{eff} (L) =-\frac12 \int d^4xd^4y \bar \psi_b \gamma_\mu
\psi_a(x) \bar \psi_{a'}(y) \gamma_\nu \psi_{b'} (y) (\delta_{aa'}
\delta_{bb'} -\frac{1}{N_c} \delta_{ab} \delta_{a'b'} ) J_{\mu\nu}
\label{s10}\ee \be \bar W (C_Q, L) =\exp \left( -\frac12
\int_{C_Q}  dx_\mu \int_{C_Q}  dy_\nu J_{\mu\nu} (x,y)\right)
\label{s11}\ee where $J_{\mu\nu}$ is
 \be J_{\mu\nu} (x,y) =\int_{L(x,x_0)} d
\Gamma_{\mu\lambda\rho} (x,z) \int_{L(y,x_0)} d\Gamma_{\nu\xi\eta}
(y,z') D_{\lambda\rho, \xi\eta} (z,z'), \label{s12}\ee with  the
gauge-invariant field correlator \be D_{\lambda\rho, \xi\eta}
(z,z') = \frac{g^2}{N_c} \lan tr F_{\lambda\rho} (z,x_0)
F_{\xi\eta} (z', x_0)\ran, \label{s13}\ee   expressed through the
parallel transported field strength, e.g. \be F_{\mu\nu} (z,x_0) =
\Phi (x_0, z) F_{\mu\nu} (z) \Phi(z,x_0), \Phi(z,x) = P\exp ig
\int^z_x A_\mu du_\mu. \label{s14}\ee
 For illustration purpose,
when only heavy quarks in $G_{QQ}$ are present, i.e.
$S_{eff}\equiv 0$ in (\ref{s9}), the point $x_0$ can be chosen
inside $C_Q$ and contours $L(x,x')$ taken as straight lines, $L\to
L_{FS}$ like in the Fock-Schwinger gauge) and $\bar W$ acquires
the standard form: \be
 \bar W (C_Q, L_{FS}) =\exp (-\frac12
\int_{S_{\min}} d\sigma_{\lambda\rho} (u) \int_{S_{\min}}
d\sigma_{\xi\eta} (v) D_{\lambda\rho, \xi\eta}
(u,v))
\label{s15}\ee
 where $d\sigma_{\lambda\rho}(u)$ is a surface
element at the point $u$, $S_{\min}$ is the minimal area surface
inside $C_Q$,  when confining part $D$  is present in
$D_{\lambda\rho, \xi\eta}$, \be D_{\lambda\rho, \xi\eta}{(u,v)} =
(\delta_{\lambda\xi}\delta_{\rho\eta} -\delta_{\lambda\eta}
\delta_{\rho\xi} ) D(u-v) + {\rm total~derivative} \label{s16}\ee
then (\ref{s15}) reduces for  large contours to the area law form,
$\bar W(C_Q, L_{FS})\cong\exp (-\sigma S_{\min})$.

Now consider the case when both terms in (\ref{s9}) are present.
To the lowest order in $S_{eff}(L)$ and integrating over light
quarks, one obtains in (\ref{s9}) a factor  $\lan
S_{eff}(L)\ran_\psi=-\frac12 \int d^4 xd^4 y  (\gamma_\mu S
(x,y)\gamma_\nu S(y,x))J_{\mu\nu}$ (S(x,y) is the light quark
propagator) similar to $G_{QQ}$, Eq.(\ref{s6}), which produces as
in (\ref{s7}),(\ref{s8}) a new Wilson loop of light quarks.  The
situation here is similar to the calculation of the vacuum average
of the product of two Wilson loops, which was done in \cite{51}.
Indeed both $\exp(-S_{int} (A))$ and $\bar W$ are linear in
$A_\mu$ (hence also in $F_{\mu\nu}$) and the general formalism
used in \cite{51} is applicable also to calculate (\ref{s9}). In
case of the connected average of two Wilson loops,
 \be
 \chi(C_1, C_2)
=\lan  W(C_1, A) W(C_2, A) \ran - \lan W(C_1, A) \ran \lan W(C_2,
A)\ran \label{s17}\ee the resulting $\chi(C_1, C_2)$ for the case
of two large contours $C_1, C_2$ in one plane with opposite
orientation  was obtained in \cite{51} for $S_2^{\min}<
S_1^{\min}$
 \be
\begin{array}{c} \chi(C_1, C_2)=\exp (-\sigma ( S_1^{\min}+ S_2^{\min }))\left\{
\frac{1}{N_c^2} \exp (2\sigma S_2^{\min})+\right.\\
\left.
 +\left(
1-\frac{1}{N^2_c}\right) \exp \left( -\frac{2\sigma
S_2^{\min}}{N_c^2-1}\right) -1\right\} \cong \frac{1}{N_c} \left[
\exp (-\sigma (S_1^{\min}-S_2^{\min}))+O\left(\frac{\sigma
S_2^{\min}}{N_c^2-1}\right) \right] \end{array}.
\label{s18}\ee

The net outcome of the analysis in \cite{51}  for a general
configuration of two oppositely oriented contours $C_1, C_2$ is
that for small distance $h$ between the surfaces $S_1^{\min} $ and
$S_2^{\min}, h<h_{crit}$ one can write \be \chi(C_1, C_2) \approx
\frac{1}{N^2_c} \exp (-\sigma S^{\min}(1,2))\label{s19}\ee where
$S^{\min}(1,2)$ is the minimal surface with the boundaries at
$C_1$ and $C_2$, and the critical distance $h_{crit}$ is defined
by the condition which can be approximately written as such
distance $h$ for which $S^{\min} (1,2) = S^{\min}_1+S_2^{\min}$.
For $h>h_{crit}$ the value of $\chi(c_1, c_2)$ is defined by
perturbative exchange of two gluons propagating in the surface
$S_{2g}^{\min}(1,2)$ (see \cite{51} for more details and
discussion) -- the two-gluon glueball exchange between contours
$C_1, C_2$.

Coming back to the calculation of (\ref{s9}) one can notice that
the same answer (\ref{s18}) for two Wilson loops can be obtained
by the proper choice of the set of contours $\{L\}$ which minimize
the common surface $S(1,2)$, namely
 \be \chi(C_1, C_2)
=\frac{1}{N_c^2}\lan \exp (-\sigma S_L (1,2))\ran_{\min
L}.\label{s20}\ee

Indeed one can show  that the contribution of the kernel
$J_{\mu\nu}$ to the product of two oppositely  oriented Wilson
loops on one plane $\bar W_1(C_Q, L) \bar W_2(C_Q,L)$ vanishes
inside the smaller loop $\bar W_2$, which yields the same answer
as in (\ref{s19}).

In a similar way the average in (\ref{s9}) can be written as \be
\begin{array}{c}
\lan e^{-S_{int(A)}}W(C_Q,A)\ran=\int D\kappa(L)
e^{-S_{eff}(L)}\bar W(C_Q,L)=\\ =\frac{1}{N_c^2} e^{-S_{eff}(L_0)}
\bar W(C_Q, L_0)\end{array}\label{s21}\ee where $L_0$ is the set
of contours which ensures the minimal area between the quark
trajectories generated by $S_{eff}$ (those will be exemplified
below) and trajectories of heavy quarks.

As a result of a light quark pair production in presence of the
heavy quark loop $W(C_Q,A)$ one thus obtains after averaging over
vacuum fields the same loop covered with the film, but with a hole
due to light  quark loop. This situation is depicted in Fig 1. Now
we turn to pion creation in the same system.

\unitlength 1mm 
\linethickness{0.4pt}
\ifx\plotpoint\undefined\newsavebox{\plotpoint}\fi 
\begin{picture}(62.25,41)(0,0)
\put(39.88,27){\oval(41.25,22.5)[]}
\put(39.88,27.5){\oval(19.25,5)[]} \put(19,27.25){\circle*{3.64}}
\put(60.5,27.25){\circle*{3.5}} \put(30,27.5){\circle*{2.24}}
\put(49,27.75){\circle*{1.5}} \put(23.75,41){Q}
\put(33.75,32.5){q} \put(34.5,22.75){q} \put(24.5,12.75){Q}
\put(34.5,10.25){Fig 1}
\end{picture}

\section{Light quarks and pions in the heavy quarkonium}

Having in mind the main result of the previous section,
Eq.(\ref{s21}), one can proceed as in the case of light quarks
without heavy quarkonium, i.e. as was done in \cite{46},
correcting in the final results for the holes, made in the world
sheet of the heavy quarkonium by light quark loops. In this way we
first obtain as in \cite{46} due to the bosonization of the
four-quark action $S_{eff}(L)$, Eq. (\ref{s10}). \be
 Z=\int D\psi  D\bar \psi
e^{-(S_1+S_{QM})}DM_sD\phi_a  d\rho (C_Q) \bar W (C_Q, L)
\label{s22}\ee where the quark-meson effective action $S_{QM}$ is
\be
\begin{array}{c}
S_{QM} =- \int d^4 x d^4y [\bar \psi^f(x)i M_s (x,y) \hat U^{fg}
(x,y) \psi^g (y)-\\ -2  N_f (J_{\mu\nu} (x,y))^{-1} M_s^2(x,y)],
~~ \hat U=\exp (i\gamma_5 t^a\phi_a
(x,y)).\end{array}\label{s23}\ee

 Integrating  out the light quarks one obtains the
Effective Chiral Lagrangian (ECL) as in \cite{46}
 \be Z= \int DM_s D\phi_a d\rho (C_Q)
\bar W (C_Q, L) e^{-S_{ECL}},\label{s24}\ee with \be S_{ECL} =
2N_f \int d^4 x d^4y J_{\mu\mu}^{-1} M_s^2(x,y) -
W(\phi),\label{s25}\ee where \be W(\phi) =N_c tr\ln [ i (\hat
\partial + \hat m+ M_s(x,y) \hat U)]
\label{s26}\ee
The integration over $DM_s D\phi_a$ is done in a standard way
using the stationary point equations in (\ref{s24}) which yields
\be
\phi^{(0)}_a=0,~~ M_s^{(0)} (x,y) =\frac{N_c}{4N_f} J_{\mu\mu}
(x,y) Tr (S(x,y))
\label{s27}\ee
 where $S(x,y) =S_\phi
(x,y)|_{\phi=0}$, and \be S_\phi(x,y) =\lan x| (i\hat \partial +
i\hat m + i M_s^{(0)} \hat U)^{-1}|y\ran. \label{s28}\ee

In what follows we shall be interested in heavy quarkonia decays
via the OZI allowed processes with emission of zero, one and two
(in principle, more) light mesons.

The latter are present in the factor $\hat U$ in (\ref{s26}),
which can be rewritten as \be \hat U =\exp (i \gamma_5 \phi^a t^a)
=\exp \left(i \gamma_5 \frac{\varphi_a \lambda_a}{f_\pi}\right),~~
\varphi_a\lambda_a \equiv \sqrt{2} \left(\begin{array}{lll}
\frac{\eta}{\sqrt{6}}+\frac{\pi^0}{\sqrt{2}},& \pi^+,& K^+\\
\pi^-,&\frac{\eta}{\sqrt{6}}-\frac{\pi^0}{\sqrt{2}},& K^0\\ K^-,&
\bar K_0, &-\frac{2\eta}{\sqrt{6}}\end{array}\right)\label{s29}\ee
and $f_\pi\equiv 93$ MeV.

Moreover the asymptotic solution of stationary point  equations
(\ref{s27}), performed in \cite{52} yields
$(|\vex-\vex_0|\gg\lambda$)
 \be M_s^{(0)} (x,y) \approx
\sigma |\vex -\vex_0(L)| \delta^{(4)}(x-y) \equiv M(\vex)
\delta^{(4)} (x-y) \equiv M(\vex) \delta^{(4)} (x-y)
\label{s30}\ee where $\vex_0(L)$ is defined by the choice of set
of contours $L$, and as we argued  above, the minimal set of
contours corresponds to the minimal area of the surface between
heavy  quark trajectories with the hole made by light quark loops
made of $S_\phi(x,y)$. One can easily understand that  this
situation is realized when $\vex_0(L)$ for a given light quark is
chosen at the position of the heavy antiquark, and correspondingly
for the light antiquark.

One important conclusion from (\ref{s30}) is that  for large
distances $R\gg\lambda$, one can use only local quantities for
$M_s^{(0)}$ and $\varphi_a,\varphi_a (x,y)\to \varphi_a(x)$.

As the next step we expand $W(\phi)$ in powers of the Nambu-Goto
fields $\phi_a$ and consider separately terms of the zeroth, first
and second power in $\phi_a$  in the partion function $Z$.

One has \be \begin{array}{c} W(\phi)\cong N_c tr \ln [ S^{-1} -
M\gamma_5 \hat \phi-\frac{i}{2} M\hat \phi^2]=\\ =W_0(\phi) +
W_1(\phi) +W_2(\phi) +..., \hat \phi\equiv t^a
\phi^a=\frac{\varphi_a\lambda_a}{f_\pi},\end{array}\label{s32}\ee
\be W_2(\phi) =-\frac{N_c}{2} tr (iSM\hat \phi^2+ SM\hat \phi
\gamma_5 S\gamma_5 M\hat \phi).\label{s33}\ee

In an analogous way one obtains terms of higher order in $\hat
\phi$, which correspond to the decay processes of heavy quarkonia
with emission of three and more $NG$ mesons.

At this point one can realize, that the interaction between heavy
quark loop and light quark (and pions) is mediated by $M_s(x,y)$,
which is string-like for large distances (large loops), as it is
in physical situation. For small loops, one has instead in
(\ref{s12}) ($M_s$ is proportional to $J_{\mu\nu})$ gluonic
condensate $\lan F^2_{\mu\nu}(x)\ran$, as it is in \cite{14,31}.
Below we shall discuss the large loop situation, and to this end
we need  to detalize the hadronic states in the big heavy quark
loop with the hole made by light quark loop.

Consider now the two-pion term $W_2(\phi)$ in the  background of
the heavy quark loop $W(C_Q,L)$, as  in (\ref{s24}).

The amplitude is  proportional to \be \lan  W_2(\phi)\ran_Q\equiv
\int  d \rho (C_q) W_2(\phi) W(C_Q, L)\label{33}\ee One can
rewrite $W_2(\phi)$ using \cite{53} as follows \be W_2(\phi)
=\frac12 \int \frac{d^4k_1}{(2\pi)^4} \frac{d^4k_2}{(2\pi)^4}
\phi_a (k_1) N(k_1, k_2) \phi_a (k_2)\label{34}\ee with
 $$
N(k_1, k_2) =\frac{N_c}{2} \left\{ \int dx e^{i(k_1+k_2)x}tr
(\Lambda M_s)_{xx}+\right.$$  \be\left.
 +\int d^4 x d^4y e^{ik_1x+ ik_2y} tr
(\Lambda(x,y) M_s(y) \bar \Lambda (y, x)
M_s(x))\right\}\label{35b}\ee

and \be \Lambda= (\hat \partial + m+ M_s)^{-1}, ~~ \bar \Lambda
=(\hat \partial - m- M_s)^{-1}.\label{36b}\ee

Both terms in (\ref{35b}) are depicted in Fig.2, the first-tadpole
term in Fig. 2a, and the second-two-point quark loop in Fig. 2b.

\vspace{1cm}

\unitlength 1mm 
\hspace{-1cm}
 \linethickness{0.4pt}
\ifx\plotpoint\undefined\newsavebox{\plotpoint}\fi 
\begin{picture}(125.5,40)(0,0)
\put(43.13,28.75){\oval(44.25,22.5)[]}
\put(103.63,28.88){\oval(43.75,21.25)[]}
\put(43.5,29.63){\oval(20,5.75)[]}
\put(104.38,30.5){\oval(19.75,6)[]} \put(33.5,30){\circle*{3}}
\put(94.5,31.5){\circle*{.5}} \put(94.5,30.5){\circle*{2.06}}
\put(114,31.25){\circle*{2.06}}
\multiput(25.68,33.93)(.064103,-.032051){13}{\line(1,0){.064103}}
\multiput(27.35,33.1)(.064103,-.032051){13}{\line(1,0){.064103}}
\multiput(29.01,32.26)(.064103,-.032051){13}{\line(1,0){.064103}}
\multiput(30.68,31.43)(.064103,-.032051){13}{\line(1,0){.064103}}
\multiput(32.35,30.6)(.064103,-.032051){13}{\line(1,0){.064103}}
\multiput(24.93,23.68)(.040509,.033565){18}{\line(1,0){.040509}}
\multiput(26.39,24.89)(.040509,.033565){18}{\line(1,0){.040509}}
\multiput(27.85,26.1)(.040509,.033565){18}{\line(1,0){.040509}}
\multiput(29.3,27.3)(.040509,.033565){18}{\line(1,0){.040509}}
\multiput(30.76,28.51)(.040509,.033565){18}{\line(1,0){.040509}}
\multiput(32.22,29.72)(.040509,.033565){18}{\line(1,0){.040509}}
\put(84.93,30.93){\line(1,0){.925}}
\put(86.78,30.88){\line(1,0){.925}}
\put(88.63,30.83){\line(1,0){.925}}
\put(90.48,30.78){\line(1,0){.925}}
\put(92.33,30.73){\line(1,0){.925}}
\multiput(94.18,30.68)(-.0625,.0313){4}{\line(-1,0){.0625}}
\put(113.68,31.18){\line(1,0){.972}}
\put(115.62,31.18){\line(1,0){.972}}
\put(117.57,31.18){\line(1,0){.972}}
\put(119.51,31.18){\line(1,0){.972}}
\put(121.46,31.18){\line(1,0){.972}}
\put(38,9.75){Fig 2 (a)} \put(96,10.75){Fig 2 (b)}
\end{picture}

The resulting dipion spectrum is defined by $N(k, k')$ averaged
with the heavy quark Green's function (only one part of it -- the
Wilson loop $W(C_Q,L)$ is present in (\ref{33}) for  simplicity of
discussion), so that one writes for the dipion spectrum with the
phase space factor $d\Phi$ \be d\Gamma_{\pi\pi} \sim |\lan N(k_1,
k_2)\ran_Q|^2 d\Phi_{(X\pi\pi)}.\label{37b}\ee

At this point it is convenient to discuss the PCAC limit  of our
expressions (\ref{33}), (\ref{35b}), e.g. the so-called Adler zero
problem. In terms of $N(k,k')$ this amounts to the requirement of
vanishing of $\lan N(0,0)\ran_0$ in the chiral limit, $m_q=m\to
0$.

In the vacuum average case, when no heavy quarks are present, this
property was proven in \cite{53}.

Indeed, one can write  $$\lan N(0,0)\ran_0 =\frac12 \lan tr \{
(\Lambda M_s)_0 + \int d^4 z e^{ikz}\Lambda(0,z) M_s (z) \bar
\Lambda (z,0) M_s(0)\}\ran_0=$$\be =\frac12 \lan tr (\Lambda M_s
\bar\Lambda (\hat\partial - m))\ran_0=\frac12 m tr \Lambda =
-\frac{m}{2N_c} \lan tr \bar \psi \psi\ran =\frac{m^2_\pi
f^2_\pi}{4N_c}.\label{38b}\ee

We have used here identical transformation for the quark loop:
$$ \lan tr (\Lambda M_s)\ran_0 = \lan tr
(M_s\Lambda\bar\Lambda^{-1}\bar\Lambda)\ran_0= \lan tr
(M_s\gamma_5\Lambda\gamma_5 (M_s+m)\Lambda)\ran_0$$ exploiting the
symmetry $x_\mu\to-x_\mu$ of the vacuum.

Thus in pure vacuum one obtains expansion of  $N(k,k') = O(m)+
O(k^2, k^{'2}, kk')$. Note the  relative  negative  sign of two
contributions in Fig.2a and Fig.2b However, in the  realistic
situation of the reaction $X(n)\to X(n')\pi\pi$ the lowest order
terms $O(k_{1\mu} k_{2\nu})$ from the expansion of $N(k_1, k_2)$
are multiplied by (among others) momenta of heavy quarks, so that
a typical term in the process amplitude would be $\frac{
(\mathcal{P}k_1)( \mathcal{P}k_2)}{\mathcal{P}^2}$, where
$\mathcal{P}$ is the momentum of $X(n)$. In the c.m. system it
amounts to amplitude $\sim \omega_1  \omega_2$, which according to
Eq. (\ref{a1'.6}) is $\sim \frac{(\Delta M)^2}{4}$, where $\Delta
M$ is the mass difference of $X(n)$ and $X(n')$.  This factor does
not provide any noticeable damping of $\pi\pi$ spectrum near
threshold. Therefore we do not have any reason to believe that the
Adler zero argument alone can play any role in the building up
the $\pi\pi$ spectrum. An additional support of this statement
comes from experiment, where spectra in $\Upsilon(3S)\to \Upsilon
(1S)\pi\pi, ~~ \Upsilon (2S)\to \Upsilon(1S)\pi\pi$ and
$\Upsilon(3S)\to \Upsilon(2S)\pi\pi$ do not have a universal
damping near $\pi\pi$ threshold \cite{42}, but are  quite
different. At the same time the spectrum of $\psi(2S) \to
J/\psi(1S)\pi\pi$ \cite{54} behaves like $\sim (q^2-4m^2_\pi)^2$,
as suggested in \cite{55}. This behaviour is not deducible from
Adler zero arguments.

However, the phenomenon of two terms with different signs in
Fig.2a and 2b, fully compensating each other in the chiral limit,
is vitally  important for our final results, where these two
amplitudes will have destructive interference. Moreover, in
section 7 we show that applying the Adler zero requirement to our
final equation, one predict the amplitude of $\pi\pi$ production
in good agreement with experiment.

\section{Derivation of the  decay transition}

The original idea of the mechanism discussed in this section was
given in \cite{45}, and was called there the Chiral Decay
Mechanism (CDM).

To generate the string-breaking vertex one can use the quark-meson
Lagragian  (\ref{s23}), which for the mesonless decays can be
written as the  string-breaking vertex operator \be
\begin{array}{l} S_{QM} \to S_{SBr} = -i \int d^4 x \int d^4 y \bar
\psi (x) M_s (x,y) \hat U(x,y)\psi (y) \to\\ \to -i \int d^4 x
\bar \psi (x) M_{br}  \hat U(x)\psi(x).\end{array} \label{s34}\ee
We have introduced in (\ref{s34}) the local  limit of the mass
operator $M_s(x,y)$, which we call the vertex mass operator
$M_{br}$. It differs from the usual  mass operator
$M_s(x)=\sigma|\ver|$ placed at the end of the string and
incorporating confinement, as exemplified in (\ref{s30}), because
$M_{br}$ is positioned somewhere in the middle of the string and
at the beginning (or the end point) of the light quark loop.
Therefore the calculation of $M_{br}$ needs a special care, and
was originally done  approximately in \cite{45}. One should stress
that  $M_{br}$  is placed at the pion emission vertex of the
quark-loop hole in the  film of $Q\bar Q$ string, a similar
important role is playing by  the vertex mass $M(0)$ in the
current correlator diagram  in \cite{53} and the value $M(0)\sim
 150$ MeV for $u,d$ quarks was used to calculate
$f_\pi$ and $\lan \bar q q\ran$.

However dynamics in $M_{br}$ and $M(0)$ is different, since the
former defines the decay process of a long string, while the
latter refers to the amplitude of $q\bar q$ meeting together at
one point, with the (short) string between $q$ and $\bar q$.
Therefore $M(0)\sim \sigma \lambda$, while $M_{br}$ is expected to
be of the order of $\lan m+\hat D\ran\sim m_q+\bar \omega$. In
what follows we shall  consider $M_{br}$ as a free and universal
parameter, and we shall  fix it once from the  decay
$\psi(3770)\to D\bar D$ and use for all dipion decays of
$\Upsilon$ and $\psi$ families. As we shall see, this strategy
gives a good result for the absolute values $\Gamma_{\pi\pi}$, at
least for $\Upsilon(2S)\to \Upsilon (1S)\pi\pi$.

To proceed  we shall define the problem of quarkonia decay with
emission of a  light quark pair and any  number of NG mesons. To
work in a relativistic invariant way we need to define the
vertices (currents) for creation of heavy quark pair in a given
state at point 1, $ J_{QQ}^\Gamma= \bar \Psi_Q \Gamma_1 \Psi_Q$,
with $\Gamma_1=\gamma_i, \gamma_5, \gamma_5\gamma_i$ etc. and the
vertex for the creation at the point $x$ of a light quark pair
plus possibly some number of NG mesons is given in (\ref{s34}),
$\Gamma_x=M_{br} \hat U(x)$. It is clear, that the general outcome
of this creation will be production of two heavy-light mesons at
points 2 and 3 with vertices $\bar \psi_Q \Gamma_2 \psi_q$ and
$\bar \psi_q\Gamma_3 \Psi_Q$, see Fig.3. The connected Green's
function for such production, $G_{123x}$, is expressed via quark
Green's functions $S_{Q(q)} (i,k)$ and has the form
$$
G_{123x} =\lan 0 | j_1j_2j_3j(x)|0\ran=$$ \be = \lan tr (\Gamma_1
S_Q (1,2) \Gamma_2 S_q (2, x) \Gamma_x S_q (x,3) \Gamma_3 S_q
(3,1))\ran.\label{40z}\ee

\unitlength 1mm 
\hspace{2cm}
 \linethickness{0.4pt}
\ifx\plotpoint\undefined\newsavebox{\plotpoint}\fi 
\begin{picture}(61,42)(0,0)
\multiput(16,22.75)(.03373016,.05753968){126}{\line(0,1){.05753968}}
\multiput(20.25,30)(.03605769,.03365385){104}{\line(1,0){.03605769}}
\multiput(24,33.5)(.0485075,.0335821){67}{\line(1,0){.0485075}}
\multiput(27.25,35.75)(.0721154,.0336538){52}{\line(1,0){.0721154}}
\multiput(31,37.5)(.1222222,.0333333){45}{\line(1,0){.1222222}}
\multiput(36.5,39)(.316667,.033333){30}{\line(1,0){.316667}}
\multiput(46,40)(.46875,-.03125){8}{\line(1,0){.46875}}
\multiput(49.5,39.75)(-.04365079,-.03373016){126}{\line(-1,0){.04365079}}
\multiput(44,35.5)(-.0634328,-.0335821){67}{\line(-1,0){.0634328}}
\multiput(39.75,33.25)(-.051346801,-.033670034){297}{\line(-1,0){.051346801}}
\put(15.75,22.75){\line(4,-5){6}}
\multiput(21.75,15.25)(.0583333,-.0333333){60}{\line(1,0){.0583333}}
\multiput(25.25,13.25)(.1111111,-.0333333){45}{\line(1,0){.1111111}}
\multiput(30.25,11.75)(.402174,-.032609){23}{\line(1,0){.402174}}
\put(39.5,11){\line(1,0){10.75}}
\multiput(50.25,11)(-.05851064,.03368794){141}{\line(-1,0){.05851064}}
\multiput(42,15.75)(-.0758427,.0337079){89}{\line(-1,0){.0758427}}
\put(35.25,18.75){\line(-2,1){10.5}}
\put(16.25,22.75){\circle*{4.03}} \put(48.75,39.75){\circle*{4.5}}
\put(25.25,24){\circle*{.5}} \put(25.75,23.5){\circle*{4}}
\put(49.75,11.25){\circle*{4.74}} \put(10.5,23){1}
\put(59.25,40.75){2} \put(61,10.75){3} \put(31.5,24){x}
\put(28,5){Fig 3}
\end{picture}

As it  is shown in Appendix 1 and 2, one can use  for $S_{Q(q})$
the Fock-Feynman-Schwinger Representation (FFSR), see \cite{56}
for review), where all Dirac bispinor structure is factorized in
the first approximation (neglecting spin-dependent forces as
compared to confinement potential). Thus one introduces the
factors $\bar Z$ (see Eq. (\ref{A.11}) in Appendix 1),  and the
rest part of $G_{123x}$ does not contain spin factors, but rather
overlap integrals of the corresponding scalar wave functions, when
the spectral representation of Green's functions is used, see Eq.
(\ref{a1'.21}) in Appendix 2.

The remaining problem is now to go over from the current
correlators, i.e. 4 point Green's function $G_{123x}$, to
hadron-hadron correlators, which is done by amputating the
point-to-hadron pieces, proportional to the decay constants of
given currents, as shown in Appendix 2.

The corresponding ``point-to-hadron'' matrix element is given by
(see (\ref{a1'.11}) for details)\be \lan 0|j_\Gamma|n, \veP\ran  =
\varepsilon_\Gamma\sqrt{\frac{E_n}{2}} f_\Gamma^{(n)},~~
\varepsilon_\Gamma=1~~{\rm for~~scalars},~~
\varepsilon_V=\varepsilon_\mu ~~{\rm for~~vectors}\label{41z}\ee
where decay constants $f^{(n)}_\Gamma$ are computed through
solution $\varphi_n(r)$ of relativistic Hamiltonian, as in
\cite{57}, \be |f^{(n)}_\Gamma|^2=\frac{6|\varphi_n(0)|^2\bar
Y_\Gamma}{M_n},\label{42z}\ee and \be \bar
Y_{\Gamma_1\Gamma_2}=\frac{tr(\Gamma_1(m_1-\hat
D_1)\Gamma_2(m_2-\hat D_2))}{4\bar \omega_1, \bar
\omega_2},\label{43z}\ee i.e. the same as the factor $\bar{ Z}$
for the two-point function, $\bar Y_{\Gamma_i\Gamma_i} =\bar Z_i$,
see Appendices 1,2 for details.

Now dividing $G_{123x}$ by matrix elements (\ref{41z}) and taking
Fourier transform in the coordinates of points 1,2,3, one obtains
the hadron decay amplitude, as derived in
(\ref{a1'.23})\footnote{a similar procedure is applied in lattice
simulations, e.g. in case of semileptonic decay formfactors, see
\cite{*}} \be\lan n_1\veP_1 | G| n_2 \veP_2,  n_3 \veP_3\ran\equiv
\frac{(\int G_{123x} d^4x)_{\veP_1,\veP_2,\veP_3}}{\Pi\lan
0|j_i|n_i\veP_i\ran \exp (-\sum_i E_{n_i}t_i)}=\label{44z}\ee
$$
=\frac{M_{br}}{\sqrt{N_c}} (2\pi)^4 \delta^{(4)}
(\mathcal{P}_1-\mathcal{P}_2-\mathcal{P}_3) J_{n_1n_2n_3} (\vep)$$
and defining $\bar y_{123} =\frac{\bar Z_{123x}}{M_{br}
\sqrt{\prod^3_{i=1}\bar Z_i}}$, one can write \be J_{n_1n_2n_3}
(\vep)= \bar y_{123} \int d^3 (\vev-\veu) d^3(\vex-\veu)
e^{i\vep\ver} \psi^*_{n_1} (\veu-\vev) \psi_{n_2} (\veu-\vex)
\psi_{n_3}(\vex-\vev)\label{45z}\ee where $\psi_{n_i}$ are
solutions of the corresponding Hamiltonians for heavy-heavy
$(n_1)$ and heavy-light quarkonia, see Appendix 1,2, and Eq.
(\ref{A.14}) for details.

To proceed  we introduce  $Q\bar Q$ Green's functions, in the
energy representation with the c.m. momentum equal to zero, having
spectral representation  (without light quarks)\be G_{Q\bar Q}
(\ver, \ver', E) =\sum_n \frac{\psi^{(n)}_{Q\bar Q}
(\ver)\psi^{(n)+}_{Q\bar Q} (\ver')}{E-E_n}\label{41}\ee and for
the $Q\bar Q$ Green's function with insertion of the light quark
loop one has \be G^{q\bar q}_{Q\bar Q} (1, 2; E) =\sum_{n,m}
\frac{\psi^{(n)}_{Q\bar Q}(1) w_{nm}(E) \psi^{(m)+}_{Q\bar
Q}(2)}{(E-E_n) (E-E_m)}\label{42}\ee where
 \be w_{nm}(E) = \gamma
\int \frac{d^3\vep}{(2\pi)^3} \sum_{n_2n_3} \frac{J_{nn_2n_3}
J^+_{mn_2n_3}}{E-E_{n_2n_3}(\vep)},\label{43}\ee
  Here $
\gamma=\frac{M_{br}^2}{N_c}$ and factors $\bar Z$ are computed in
the Appendix 1 together with details of derivation and definition
of einbein masses $\omega_k, \Omega_k$.

It is easy to derive from (\ref{43}) an expression for the
(complex ) shift of the n-th energy level, which has the same form
as in the general multichannel theory (see \cite{58} for a review,
and \cite{59,60} for a recent application to the shift of $D^*_s,
B_s^*$ masses). \be E-E_n=- w_{nn}(E)\label{46}\ee and the width
in the channel $k$ is obtained as (for two equal mass final
heavy-light mesons) \be \Gamma_n = \gamma \frac{p_k\tilde
M_k}{4\pi^2} \int d\Omega_{\vep}|J_{nn_2n_3}(\vep)|^2\label{47}\ee
where $\tilde M,~ p_k$ is the reduced mass and relative momentum
in the channel $k$.

Expressions (\ref{46}),(\ref{47}) are sufficient to obtain shift
and width of any state due to a light pair creation, provided the
dynamics of heavy-light mesons is known (wave functions and
einbein masses (average energies)).  As was pointed out before,
additional NG meson emission is described by simply multiplying
integrands of matrix elements $J_{nn_2n_3}$ with terms of
expansion $\hat U(x) = \exp \left(i\gamma_5
\frac{\varphi_a\lambda_a}{f_\pi}\right)=1+ i\gamma_5
\frac{\varphi_a\lambda_a}{f_\pi} +..., $ $
\varphi_a=\frac{e^{i\vek\vex}}{\sqrt{2\omega(\vek)V_3}}$, adding
meson energy in the denominator and modifying $\ver$ in
(\ref{45z}). In the next section we discuss   as a clarifying
example the  decay without meson emission , e.g. $\psi(3770)\to
D\bar D$  and in section 6 emission of two pions.

\section{Heavy quarkonia decays without pion emission}

In this section we calculate explicitly the operator $W(E)$ in
(\ref{42}), (\ref{43}), which reduces to the calculation of the
transition  matrix element $J_{n_1n_2n_3}$  (\ref{45z}). As a
practical example we have in mind the  decay $\Upsilon (4S)\to B
\bar B$, or $\psi(3770)\to D\bar D$, however our results will be
applicable to other two-body decays of this sort.  We start with
the calculation of the factor $\bar Z_{123x}$, given in
(\ref{45z}) and Appendix 1, where we should insert
$\Gamma_Q=\gamma_i, \Gamma_{Q\bar q}=\gamma_5=\Gamma_{q\bar Q},
\Gamma_x=1$. The resulting trace is easily computed and can be
written as $(Z_{123x} \equiv Z)$
$$ Z=tr_L[\gamma_i(m_Q+\Omega_Q\gamma_4-i\hat p_Q)\gamma_5(m_{\bar
q}-\omega_{\bar q}\gamma_4 +i\hat p_{\bar q})\times$$\be
\times(m_q+\omega_q\gamma_4-i\hat p_q)\gamma_5(m_{\bar Q}-
\Omega_{\bar Q}\gamma_4+i\hat p_{\bar Q})].\label{5.1} \ee Here
notations are $\hat p_k=\sum^3_{i=1} p^i_k\gamma^i$. For the
$B\bar B$ or $D \bar D$ decays in the c.m.system  one  can
simplify $ m_Q= m_{\bar Q}, \Omega_Q=\Omega_{\bar Q},$ and one
finally has
$$ Z=4\{i2m_Q \Omega\omega (p_{qi} -p_{\bar q i}) + i \omega^2
m_Q (p_{Qi}-p_{\bar Q i})$$ \be + im (p_{Qi} (p_q p_{\bar q}) +
p_{qi} (p_{\bar Q} p_{\bar q}) -p_{\bar qi} (p_{\bar Q} p_q) -
p_{\bar Qi}(p_{q}p_{\bar q})-  p_{\bar qi} (p_Qp_q) + p_{qi} (p_Q
p_{\bar Q} ) )\}\label{5.2}\ee

One can define relative momenta in the $Q\bar q$ and $\bar Q q$
systems so that in the total c.m. system  (see Appendix 3 for
details) \be \frac{\vep_Q-\vep_{\bar q}}{2}=\veq_1,~~
\frac{\vep_{\bar Q}-\vep_{\bar q}}{2}=\veq_2,~~ \vep_Q +\vep_{\bar
q} =\vep, ~~\vep_{\bar Q} +\vep_{ q} =-\vep, \label{5.3}\ee and
finally \be \vep_Q =\veq_1 + \frac{\vep}{2};~~  \vep_{\bar q} =-
\veq_1 +\frac{\vep}{2};~~ \vep_{\bar Q} =\veq_2 -\frac{\vep}{2};~~
\vep_q=-\veq_2 -\frac{\vep}{2}.\label{5.4}\ee

The  normalized factor  $\bar Z =
\frac{Z}{\Pi_{k=q,Q}(2\omega_k)}$ is for large mass $m_Q$ \be \bar
Z\cong \frac{4\cdot  2m_Q i \Omega\omega ( q_{1i} -
q_{2i}-p_i)}{16\Omega^2 \omega^2} = \frac{im_Q(q_{1i} - q_{2i}
-p_i)}{2\Omega \omega}.\label{5.5}\ee

The transition matrix element $ J_{n_1n_2n_3}(\vep) \equiv
J(\vep)$ is \be J(\vep) = \int \bar Z e^{i\vep \ver} d^3(\veu-
\vev) d^3 (\vex -\veu) \Psi^+_{n} (\veu-\vev)
\psi_{n_2}(\veu-\vex) \psi_{n_3} (\vex-\vev)\label{5.6}\ee where
$\ver = \frac{\Omega (\veu-\vev)}{\omega+ \Omega} \equiv
c(\veu-\vev)$.

It is convenient to introduce Fourier transform of wave functions,
\be \Psi_{n} (\vez) =\int \frac{d^3q' e^{i \veq'\vez}}{(2\pi)^3}
\tilde \Psi_{n} (\veq'), ~~\psi_{n_2} (\veu-\vex) = \int
e^{i\veq_1 (\veu-\vex)} \frac{d^3\veq_1}{(2\pi)^3} \tilde
\psi_{n_2}(\veq_1)\label{5.7}\ee
$$ \psi_{n_3} (\vex -\vev) = \int \frac{d^3\veq_2}{(2\pi)^3}
e^{i\veq_2(\vex-\vev)}\tilde \psi_{n_3} (\veq_2).$$ As a result
one obtains for $J(\vep)$ \be J(\vep) = \int\bar y_{123} \frac{
d^3\veq_1}{(2\pi)^3} \tilde \Psi_{n} ( c\vep + \veq_1) \tilde
\psi_{n_2}(\veq_1) \tilde \psi_{n_3}(\veq_1)\label{5.8}\ee and
$\bar Z$ simplifies,
 \be \bar y_{123}= \frac{\bar Z (\veq_1 =\veq_2, \vep)}{\sqrt{\bar
 Z_1\bar Z_2\bar Z_3}};~~
   \bar Z (\veq_1
=\veq_2, \vep) \cong - \frac{im_Q p_i}{2\Omega\omega} +O\left(
\frac{p}{m_Q}\right).\label{5.9}\ee As one can see from
(\ref{5.9}) , $\bar Z$ is proportional to $p_i$ which signals the
$P$ wave of relative $D\bar D$ or $B\bar B$ motion, as it should
be.

We turn now to the calculation of the factor $y_{123}$ in
(\ref{45z}), which can be written (since $M_{br}$ is factored out
in $\Gamma_x$ and $Z$, Eq. (\ref{5.1})).

\be \bar y_{123} =\frac{\bar Z}{\sqrt{\bar Z_1\bar Z_2\bar
Z_3}}\label{60}\ee with $\bar Z$ given in  (\ref{5.9}) and $\bar
Z_i$ computed in \cite{57},  $$\bar Z_i =\frac{tr (\Gamma_i(m_1- i
\hat p_1) \Gamma_i (m_2 + i\hat p_2))}{4\bar
\omega_1^{(i)}\bar\omega_2^{(i)}}=$$\be= \frac{m_1m_2+\bar
\omega_1\bar\omega_2 +\eta_i\lan
\vep^2\ran}{\bar\omega_1\bar\omega_2}.\label{61}\ee

Here $\eta_i=+\frac13$ for the vector and -$L$ for the
pseudoscalar case. Taking into account, that with the reasonable
accuracy $\bar \omega_i\cong\sqrt{\lan \vep^2\ran + m^2_1}$, one
can compute $\bar Z_i$ for the vector $(\psi(nS), \Upsilon(nS))$
the scalar ($D,B)$ cases, \be \bar Z_D=
\frac{(\Omega-\omega)}{\Omega} ,~~\bar Z_\psi =\frac{\frac43
\Omega^2_\psi +\frac23 m^2_c}{\Omega^2_\psi}\label{62}\ee where we
have denoted  $\Omega\equiv \Omega_D$ in $D, \omega\equiv
\omega_q, q=u,d$ and approximately (see Tables in Appendix 1)
$\Omega_D =\Omega_\psi\approx 1.5$ GeV for $m_c=1.4$ GeV.

As result $\bar y_{123} (\psi \to D \bar D) \approx \bar Z \cdot
1.1=1.1 \frac{ (-ip_i)m_c}{2\omega\Omega}.$

To compare the decay probability of $\psi(3770)$ one can use
(\ref{47}) to write (we use $\lan p^2_i\ran =\frac13 \lan
\vep^2\ran)$ \be \Gamma(\psi\to D\bar D) = \left(
\frac{M_{br}}{2\omega}\right)^2\frac{p^3M_D}{6\pi N_c}\left(
\frac{1.1 m_c}{\Omega}\right)^2 |J(\vep)|^2.\label{63}\ee

One can see the correct $p^3$ behaviour of the $p$-wave decay, and
the $\frac{1}{N_c}$ factor for the string breaking effect, as it
should be.

At this point one should use realistic  wave functions of $\psi,D$
to compute $J(\vep)$. Here and below in the paper we exploit for
semiquantitative estimates the SHO wave functions, $\psi_i(p) =
\mathcal{P}_i (\vep)\exp \left(
-\frac{\vep^2}{2\beta^2_i}\right)$, given in Appendix 4 together
with matrix elements, where oscillator parameters are fitted to
reproduce the known r.m.s. radii of states, $ \lan r^2\ran_\psi =
(0.8$ fm)$^2$,   $\lan r^2\ran_D = (0.57$ fm$)^2$ \cite{61}, which
yields $\beta_\psi \equiv \beta_1 = 0.467$ GeV,  $ \beta_D\equiv
\beta_2 =0.42$ GeV and $J(p)$ takes the form \be J^{(SHO)}{(\vep)}
=(2\sqrt{\pi})^{3/2}\sqrt{\frac32}\left(
\frac{2\beta_1}{\Delta}\right)^{3/2}\left( y-\frac83
\frac{\beta_1^2(c\vep)^2}{\Delta^2}\right)
e^{-\frac{c^2\vep^2}{\Delta}}\label{64}\ee with $\Delta=
2\beta^2_1 +\beta^2_2,~~ -y
=\frac{\beta^2_2-2\beta^2_1}{\Delta},~~
c=\frac{\Omega}{\omega+\Omega}$.

For the reaction  $\psi(3770)\to D\bar D$ one has $p=0.276$ GeV
and  $J^{(SHO)}(\vep) =5.09$(GeV)$^{-3/2}$. As the result one
obtains \be \Gamma=19{\rm~ MeV~~}\left(
\frac{M_{br}}{2\omega}\right)^2\left(\frac{J}{5.09{\rm~
GeV}^{-3/2}}\right)^2.\label{65}\ee

Since in PDG $\Gamma_{exp}=23.6$ MeV one expects in the SHO
approximation, that \be M_{br}\cong 2\omega\approx
1{\rm~GeV}.\label{66}\ee

This estimate will be used below for dipionic transitions. (Note,
however, that our treatment of $\psi(3770)$ as the $2S~~ c\bar c$
state is a crude approximation, $\psi(3770)$ being mixture of
$~^3D_1, ~^3S_1$ states, therefore a mixing coefficient  should
appear as a factor of $J$. In any case our expression (\ref{65})
is rather qualitative, giving an order of magnitude estimate of
the decay process).

Exact calculation is sensitive to the details of the wave function
and will be given in a separate publication. One can say, that
calculation of $\Gamma$ is a good check of the form of  wave
functions, which can be used in addition to radiative transitions
and lepton decay width calculations, checking it with experiment.
As it is, we can now turn to the  main subject of paper, having a
formalism without fitting parameters.

\section{ Heavy quarkonia decays with emission of two NG mesons}

We shall consider in this section the string breaking in  heavy
quarkonia with emission of two NG mesons. Here we come back to
Eq.(\ref{s23}) for the string breaking action and using it one has

\be S_{SBr} = -i \int d^\eta x \bar \psi^f(x) M_{br} \hat
U^{fg}\psi^g(x)\label{6.1}\ee where we have  made explicit flavor
indices, and $\hat U$ is given in (\ref{s29}). In the previous
section we have considered decay without NG mesons, and
correspondingly replaced $\hat U^{fg}$ by $\delta_{fg}$ and
$M_s^{(0)}(x)$ by the vertex mass $M_{br}$. Here we take into
account higher terms of expansion of $\hat U$, namely \be \hat U
=\exp\left( i\gamma_5 \frac{\varphi_a\lambda_a}{f_\pi}\right)=\hat
1 +
i\gamma_5\frac{\varphi_a\lambda_a}{f_\pi}-\frac{\varphi_a\lambda_a\varphi_b\lambda_b}{2f_\pi^2}+...\label{6.2}\ee

We shall disregard in the first approximation interaction of
emitted NG mesons with other particles (taking FSI into account as
the next step)  and write \be
\varphi_a(x)=\frac{e^{i\vek\vex}}{\sqrt{2\omega_aV_3} },~~
\omega_a=\sqrt{\vek^2+m^2_a}.\label{6.3}\ee As before one can
define the $Q\bar Q$ Green's function containing one light quark
loop with NG mesons emitted from loop vertices at points $x$ and
$y$, which is again has the same form as in (\ref{42}), e.g. for
two pion emission one has \be G^{q\bar q,\pi\pi}_{Q\bar Q} (1;2;E)
= \sum_{n,m}\frac{\psi^{(n)}_{Q\bar Q}(1)
w^{(\pi\pi)}_{nm}(E)\psi_{Q\bar
q}^{(m)+}(2)}{(E-E_n)(E-E_m)}.\label{6.4}\ee But
$w_{nm}^{(\pi\pi)}(E)$ now consists of several terms,
corresponding to the diagrams depicted in Figs.4 and 5, namely

$$ w_{nm}^{(\pi\pi)} (E) =\gamma\left\{ \sum_k
\frac{d^3p}{(2\pi)^3}\frac{J_{nn_2n_3}^{(1)}(\vep,\vek_1)J^{*(1)}_{mn_2n_3}
(\vep,\vek_2)}{E-E_{n_2n_3}(\vep)-E_\pi(\vek_1)}+
(1\leftrightarrow 2)\right.$$

$$-\sum_{n'_2n'_3}
\frac{d^3p}{(2\pi)^3}\frac{J_{nn'_2n'_3}^{(2)}(\vep,\vek_1,\vek_2)J_{mn'_2n'_3}^*
(\vep)}{E-E_{n'_2n'_3}(\vep)-E(\vek_1,\vek_2)}-$$
\be\left.-\sum_{k^{\prime\prime}}
\frac{d^3p}{(2\pi)^3}\frac{J_{nn^{\prime\prime}_2n^{\prime\prime}_3}(\vep)J^{(2)*}_{mn^{\prime\prime}_2n^{\prime\prime}_3}
(\vep,\vek_1,\vek_2)}{E-E_{n^{\prime\prime}_2n^{\prime\prime}_3}(\vep)}\right\}\label{6.5}\ee

\unitlength 1mm 
\linethickness{0.4pt}
\ifx\plotpoint\undefined\newsavebox{\plotpoint}\fi 
\begin{picture}(120.71,44.28)(0,0)
\put(15.25,17.25){\framebox(45.5,25.5)[cc]{}}
\put(77.25,17.5){\framebox(41.75,25.75)[cc]{}}
\put(14.75,30.25){\circle*{2.69}} \put(61,31){\circle*{2.69}}
\put(77.25,31.5){\circle*{2}} \put(77,31.75){\circle*{2.5}}
\put(119.25,32){\circle*{2.92}} \put(35.5,43){\circle*{2}}
\put(39.5,43.25){\circle*{2.06}} \put(35.25,17.25){\circle*{2}}
\put(38.75,17){\circle*{2.24}} \put(99.25,43.25){\circle*{2.06}}
\put(95,43.25){\circle*{2}} \put(95,17.75){\circle*{2.5}}
\put(99.5,17.5){\circle*{2.55}}
\multiput(15.25,30.25)(.0336538,.0769231){52}{\line(0,1){.0769231}}
\multiput(17,34.25)(.0333333,.0333333){60}{\line(0,1){.0333333}}
\multiput(19,36.25)(.033333,.033333){30}{\line(0,1){.033333}}
\multiput(20,37.25)(.0444444,.0333333){45}{\line(1,0){.0444444}}
\multiput(22,38.75)(.1,.033333){30}{\line(1,0){.1}}
\multiput(25,39.75)(.125,.033333){30}{\line(1,0){.125}}
\multiput(28.75,40.75)(.1041667,.0333333){60}{\line(1,0){.1041667}}
\multiput(15.5,30.75)(.0336538,-.0817308){52}{\line(0,-1){.0817308}}
\multiput(16.5,28.5)(.0336538,-.0817308){52}{\line(0,-1){.0817308}}
\multiput(18.25,24.25)(.0333333,-.0333333){60}{\line(0,-1){.0333333}}
\multiput(20.25,22.25)(.0555556,-.0333333){45}{\line(1,0){.0555556}}
\multiput(22.75,20.75)(.1052632,-.0328947){38}{\line(1,0){.1052632}}
\multiput(26.75,19.5)(.125,-.033333){30}{\line(1,0){.125}}
\multiput(30.5,18.5)(.173913,-.032609){23}{\line(1,0){.173913}}
\multiput(38.25,17.75)(.2368421,.0328947){38}{\line(1,0){.2368421}}
\multiput(47.25,19)(.075,.0333333){60}{\line(1,0){.075}}
\multiput(51.75,21)(.0487805,.0335366){82}{\line(1,0){.0487805}}
\multiput(55.75,23.75)(.0335366,.0335366){82}{\line(0,1){.0335366}}
\multiput(58.5,26.5)(.0328947,.0657895){38}{\line(0,1){.0657895}}
\put(39,43.5){\line(3,-1){6.75}}
\put(45.75,41.25){\line(0,-1){.25}}
\multiput(45.75,41)(.1055556,-.0333333){45}{\line(1,0){.1055556}}
\multiput(50.5,39.5)(.0817308,-.0336538){52}{\line(1,0){.0817308}}
\multiput(54.75,37.75)(.0576923,-.0336538){52}{\line(1,0){.0576923}}
\multiput(57.75,36)(.0388889,-.0333333){45}{\line(1,0){.0388889}}
\multiput(59.5,34.5)(.032609,-.076087){23}{\line(0,-1){.076087}}
\multiput(77,32.25)(.03365385,-.05769231){104}{\line(0,-1){.05769231}}
\multiput(80.5,26.25)(.0336538,-.0384615){52}{\line(0,-1){.0384615}}
\multiput(82.25,24.25)(.0583333,-.0333333){60}{\line(1,0){.0583333}}
\multiput(85.75,22.25)(.0432692,-.0336538){52}{\line(1,0){.0432692}}
\put(88,20.5){\line(1,0){.5}}
\multiput(88.5,20.5)(.0855263,-.0328947){38}{\line(1,0){.0855263}}
\multiput(91.75,19.25)(.266667,-.033333){15}{\line(1,0){.266667}}
\multiput(99.5,18.25)(.0721154,.0336538){52}{\line(1,0){.0721154}}
\multiput(103.25,20)(.06443299,.03350515){97}{\line(1,0){.06443299}}
\multiput(109.5,23.25)(.04567308,.03365385){104}{\line(1,0){.04567308}}
\multiput(114.25,26.75)(.0410448,.0335821){67}{\line(1,0){.0410448}}
\multiput(117,29)(.0335821,.0597015){67}{\line(0,1){.0597015}}
\multiput(77.25,32.5)(.0333333,.0541667){60}{\line(0,1){.0541667}}
\multiput(79.25,35.75)(.03605769,.03365385){104}{\line(1,0){.03605769}}
\multiput(83,39.25)(.0666667,.0333333){60}{\line(1,0){.0666667}}
\multiput(87,41.25)(.173913,.032609){23}{\line(1,0){.173913}}
\multiput(91,42)(.0921053,.0328947){38}{\line(1,0){.0921053}}
\multiput(99.5,43.5)(.1447368,-.0328947){38}{\line(1,0){.1447368}}
\multiput(105,42.25)(.133333,-.033333){30}{\line(1,0){.133333}}
\multiput(109,41.25)(.0576923,-.0336538){52}{\line(1,0){.0576923}}
\multiput(112,39.5)(.0366667,-.0333333){75}{\line(1,0){.0366667}}
\multiput(114.75,37)(.0393258,-.0337079){89}{\line(1,0){.0393258}}
\multiput(118.25,34)(.0328947,-.0328947){38}{\line(0,-1){.0328947}}
\put(14.43,30.68){\line(1,0){.983}}
\put(16.4,30.71){\line(1,0){.983}}
\put(18.36,30.75){\line(1,0){.983}}
\put(20.33,30.78){\line(1,0){.983}}
\put(22.3,30.81){\line(1,0){.983}}
\put(24.26,30.85){\line(1,0){.983}}
\put(26.23,30.88){\line(1,0){.983}}
\put(28.2,30.91){\line(1,0){.983}}
\put(60.68,31.18){\line(1,0){.972}}
\put(62.62,31.24){\line(1,0){.972}}
\put(64.57,31.29){\line(1,0){.972}}
\put(66.51,31.35){\line(1,0){.972}}
\put(68.46,31.4){\line(1,0){.972}}
\multiput(77.18,32.43)(.12245,.03316){7}{\line(1,0){.12245}}
\multiput(78.89,32.89)(.12245,.03316){7}{\line(1,0){.12245}}
\multiput(80.61,33.36)(.12245,.03316){7}{\line(1,0){.12245}}
\multiput(82.32,33.82)(.12245,.03316){7}{\line(1,0){.12245}}
\multiput(84.04,34.29)(.12245,.03316){7}{\line(1,0){.12245}}
\multiput(85.75,34.75)(.12245,.03316){7}{\line(1,0){.12245}}
\multiput(87.47,35.22)(.12245,.03316){7}{\line(1,0){.12245}}
\multiput(77.18,31.93)(.08654,-.03269){10}{\line(1,0){.08654}}
\multiput(78.91,31.28)(.08654,-.03269){10}{\line(1,0){.08654}}
\multiput(80.64,30.62)(.08654,-.03269){10}{\line(1,0){.08654}}
\multiput(82.37,29.97)(.08654,-.03269){10}{\line(1,0){.08654}}
\multiput(84.1,29.31)(.08654,-.03269){10}{\line(1,0){.08654}}
\multiput(85.83,28.66)(.08654,-.03269){10}{\line(1,0){.08654}}
\multiput(87.56,28.01)(.08654,-.03269){10}{\line(1,0){.08654}}
\put(32.25,7){Fig 4} \put(91.25,8.5){Fig 5}
\end{picture}

Note, that in (\ref{6.5}) indices $n_in'_i$ and
$n^{\prime\prime}_i$ refer in general to three different sets of
intermediate states; also
$E(\vek_1,\vek_2)=\sqrt{\vek^2_1+m^2_\pi}+\sqrt{\vek^2_2+m^2_\pi}$.

Now $J_{nn_2n_3}(\vep)$ and $\gamma$ in (\ref{6.5}) are the same
as before and defined by (\ref{45z}) with $\bar Z_{ns}$ defined in
(\ref{A.11}) and $\Gamma_x =\hat 1$, the dash sign refers to
flavor indices). For $J^{(1)}_{nn_2n_3}(\vep,\vek_1)$  one should
take into account (\ref{6.2}) with extra factors of $
\frac{i\gamma_5\lambda_a}{f_\pi}\varphi_a$ and pion plane wave
(\ref{6.3}).

As a result one can write \be J^{(1)}_{nn_2n_3} (\vep, \vek_1) =
\int\bar Z^{(\pi)}_{nn_2n_3}\frac{e^{i\vep\ver+ i
\vek_1(\vex-\veR)}}{\sqrt{2\omega_\pi V_3}} d^3 (\veu-\vev)
d^3(\vex-\veu) \Psi^+_n(\veu-\vev)
\psi_{n_2}(\veu-\vex)\psi_{n_3}(\vex-\vev)\label{6.6}\ee where
$\bar Z^{(\pi)}_{nk}$ is now also a trace over flavor indices,
\be\Gamma_q\to \Gamma_q\frac{i\gamma_5}{f_\pi}(\varphi_a
\lambda_a)_\pi,\label{6.7}\ee and the notation
$(\varphi_a\lambda_a)_\pi$ implies the numerical matrix, obtained
from (\ref{s29}) for a given pion, e.g. for $\pi^+$ it is $
\sqrt{2}\left(\begin{array}{lll }
0&1&0\\0&0&0\\0&0&0\end{array}\right) $, while for $ \pi^0$ it is
$(\varphi_a \lambda_a)_{\pi^0}=\left(\begin{array}{lll }
0&0&0\\0&-1&0\\0&0&0\end{array}\right) $, so for $SU(2)$ isospin
group one has $(\varphi_a \lambda_a)_\pi=\pi_i \tau_i=
\pi^+\tau_-+\pi^-\tau_++\pi^0\tau_3$, while $\tau_i$ Pauli
matrices. For isospin conserving decays also one of vertices
$\Gamma_{Q\bar q}, \Gamma_{\bar Q q}$ or both, should have
nontrivial flavor structure, otherwise the flavor trace will be
nonzero only due to quark mass matrix $ \hat m=
\left(\begin{array}{lll }
m_u&0&0\\0&m_d&0\\0&0&m_s\end{array}\right) $, which enters in the
light quark Green's function (\ref{s28}). This is what happens in
the decays of the type $ \Upsilon(3S)\to\Upsilon(1S)\eta (\pi)$,
and will be considered in a separate publication.

To complete the flavor trace in $\bar Z_{nk}$ one should specify
the isospin structure of intermediate heavy-light mesons, e.g. for
the isospinors  $B\equiv \left(\begin{array}{l}B^+\\
B_0\end{array}\right)$ and  $B\equiv \left(\begin{array}{l}\bar B^0\\
B^-\end{array}\right)$ one  actually has final states in the
combination $K_B^{(\pi)}\equiv \pi_i(\bar B\tau_i B)$ and this
factor appears in $\bar Z_{nk}$ for given final state of $ B\bar
B$ (or $D\bar D$ is the case of $K_D^{(\pi)}$).

We now come to the vertex with emission of two pions, $
J^{(2)}_{nn_2n_3} (\vep,\vek_1, \vek_2)$, which is due to the
presence of the factor $\hat U$ in the  vertex, specifically of
its quadratic term, \be
-\frac{(\varphi_a\lambda_a)^2}{2f^2_\pi}=-\frac{(\pi^{(1)}_i\tau_i)
(\pi^{(2)}_k \tau_k)}{2f^2_\pi}=-\frac{1}{2f^2_\pi}
(\vepi_1\vepi_2+ ie_{ikl}\pi_{1ik} \pi_{2k}\tau_l).\label{6.8}\ee

One can notice, that in the case of one-pion vertices in the first
two terms in (\ref{6.5}), the resulting isospin structure is such
that pions appear in the total isospin $I=0$ state, unless some
charge (isospin) filtering of intermediate states is done, indeed
\be K_B^{(\pi)} K^{+(\pi)}_B=\pi_{1i}\sum_{B,\bar B}(\bar B \tau_i
B)(\bar B\tau_k B) \pi_{2k}= tr_{fl}(\pi_{1i} \tau_i \pi_{2k}
\tau_k)=2 \vepi_1 \vepi_2.\label{6.9}\ee

In contrast to that,  the $J^{(2)}_{nn_2n_3}$ term contains the
$I=1$ term, which of course vanishes for the transition between
$I=0$ states, as in $\Upsilon(ns)\to\Upsilon(n's)\pi\pi$, but can
be nonzero for $B\bar B$ final state.

 As the result, one can write $J^{(2)}_{nn_2n_3}$ as  $$J^{(2)}_{nn_2n_3}( \vep,
\vek_1, \vek_2 )= $$\be=\int\bar Z_{nn_2n_3}^{(\pi\pi)}
\frac{e^{i\vep\ver+i(\vek_1+\vek_2)\vex}
d^3(\veu-\vev)d^3(\vex-\veu)\Psi^+_n(\veu-\vev)\psi_{n_2}(\veu-\vex)\psi_{n_3}
(\vex-\vev)}{\sqrt{2\omega_\pi(\vek_1)
V_32\omega_\pi(\vek_2)V_3}}\label{6.10}\ee and $\bar
Z_{nn_2n_3}^{(\pi\pi)}$ is obtained from the general formula
(\ref{A.11}) substituting $\Gamma_x\to \left(-
\frac{\vepi_1\vepi_2}{2f^2_\pi}\right)$. In the same  way one can
consider emission of more pions.

Finally one can define the decay probability for the process
$(Q\bar Q)_{n}\to (Q\bar Q)_{n'}\pi\pi$, which is obtained from
the amplitude $w^{(\pi\pi)}_{nn'} (E)$ by the standard rules
$$
dw((n)\to (n') \pi\pi) =|w^{(\pi\pi)}_{nn'}(E)|^2\frac{V_3
d^3\vek'}{(2\pi)^3} \frac{V_3d^3\vek_2}{(2\pi)^3}\times$$ \be
\times 2\pi \delta (E(\vek_1,\vek_2)+
E_{n'}-E_{n}).\label{6.11}\ee
 It is easy to  see, that the factors $V_3$ cancel in (\ref{6.11})
 and $dw$ on the l.h.s. of (\ref{6.11}) has the dimension of mass;
 integrating over $\Pi_id^3\vek_i$ on the r.h.s. of (\ref{6.11})
 one obtains the total width in this channel
 \be \Gamma^{ (nn')}_{\pi\pi} = \int d w ((n)\to (n')
 \pi\pi).\label{6.12}\ee
Thus one obtains in (\ref{6.11}), (\ref{6.12}) the absolute rate
of the process and fixing the dipion mass \be q^2\equiv
M^2_{\pi\pi}= (k_1+k_2)^2 =
2m^2_\pi+2\omega_1\omega_2-2\vek_1\vek_2\label{6.13}\ee one
defines the dipion spectrum measured in experiment.

To proceed with the analysis we consider $J_{nn_2n_3}^{(1)}
(\vep,\vek_1)$ given by (\ref{6.6}) and to calculate the $\bar
Z_{nn_2n_3}^{(\pi)}$, one should specify the intermediate state
$n_2n_3$. It is easy to see, that the closest threshold is given
by the $S$-state of ($B\bar B^*+\bar B B^*)$ plus a pion. Denoting
by index $k$ the spin direction of $B^*$ and by  index $i$ the
spin direction of incident $(Q\bar Q)$ state, one easily
calculates $\bar Z^{(1)}_{ik}$ in the approximation, when all
momenta are small compared to the heavy mass $m_Q,
O(\frac{p}{m_Q})$. The result is (for the intermediate state
$B\bar B^*)$ \be \bar
Z^{(\pi)}_{ik}=i\frac{(\varphi_a\lambda_a)_\pi}{f_\pi}\delta_{ik}
\frac{m^2_Q+\Omega^2}{2\Omega^2}.\label{6.14}\ee Now introducing
Fourier transformed wave functions as in (\ref{5.8}), one arrives
at the final expression \be
J^{(1)}_{nn_2n_3}(\vep,\vek_1)=\frac{\bar
Z^{(\pi)}_{ik}}{\sqrt{2\omega_\pi V_3}}\int\frac{d^3q_1}{(2\pi)^3}
\tilde \Psi_{Q\bar Q} (c\vep-\frac{\vek_1}{2}+\veq_1) \tilde
\psi_{Q\bar q}(\veq_1) \tilde \psi_{\bar Q q}(\veq_1-\vek_1 ).
\label{6.15}\ee

Comparison with (\ref{5.5}) shows that $\bar Z$ in (\ref{6.15})
corresponds to the $S$-wave, and dependence on $\vek_1$ appears in
the arguments of  wave functions, which influences the resulting
spectrum of $\pi\pi$,however the resulting dependence appears to
be rather moderate.

We now turn to the term $J_{nn_2n_3}^{(2)}(\vep,\vek_1, \vek_2)$
in (\ref{6.5}), the general form of it is given in (\ref{6.10}).

First of all, we calculate $ \bar Z^{(\pi\pi)}_{nn_2n_3}$, and
define the intermediate state $B\bar B$ in the $P$ wave -- the
same as in the decay without pions. Hence the final form of $\bar
Z$ is
$$ \bar Z^{(\pi\pi)}=\left(-\frac{
\vepi_1\vepi_2}{f^2_\pi}\right)\left(-\frac{im_Q}{2\Omega\omega}\right)(p_{qi}-p_{\bar
qi})=$$\be
=\frac{i\vepi_1\vepi_2}{f^2_\pi}\frac{m_Q}{2\Omega\omega}(K_i-p_i),\veK\equiv\vek_1+\vek_2.\label{6.16}\ee
Here we have used Appendix  3 to express momenta in terms of
$\veK, \vep$, and
$\pi_1\pi_2=\pi^+_1\pi^-_2+\pi_1\pi_2^++\pi^0_1\pi^0_2$.

Finally $J^{(2)}$ can be written as \be J^{(2)}_{nn_2n_3}
(\vep,\vek_1,\vek_2)=\bar Z^{(\pi\pi)}\int \frac{d^3q_1\tilde
\Psi_{n}(c\vep+\veq_1+\frac{\veK}{2})\tilde
\psi_{n_2}(\veq_1)\tilde\psi_{n_3}(\veq_1-\veK)}{(2\pi)^3\sqrt{2\omega_\pi(k_1)V_32\omega_\pi(k_2)V_3}}.\label{6.17}\ee
Note, that in (\ref{6.5}) this matrix element is multiplied with
another one, where pions are not emitted, and it has the form
(\ref{5.8}), (\ref{5.9}) with $\Psi_{n'}$ corresponding to the
final state of heavy quarkonium.

Equations (\ref{5.8}), (\ref{6.15}) and (\ref{6.17}) give all the
necessary elements for calculation of $\pi\pi$ spectrum and total
width of two-pion transitions between states of heavy quarkonia ,
provided wave functions of initial and final $Q\bar Q$ states and
heavy-light mesons $Q\bar q$ and $\bar Q q$ are known.However, the
resulting multidimensional integrals are not easy to evaluate, and
we resort at this point again  to the SHO wave functions, as we
did it in  previous section. For the case when all wave functions
are taken as in oscillator potential,  the SHO integrals for
$J,J^{(1)}, J^{(2)}$ can be calculated in analytic form.

Indeed, using Fourier transforms of oscillator wave functions as
in (\ref{5.8}), (\ref{6.15}), (\ref{6.17}) and writing \be \tilde
\Psi_{n}(q,\beta)=\mathcal{P}_n (q)
e^{-\frac{q^2}{2\beta^2}}\label{6.18}\ee where
$\mathcal{P}_1(q)=\left(\frac{2\sqrt{\pi}}{\beta}\right)^{3/2}$,
$\mathcal{P}_2(q)=\left(\frac{2\sqrt{\pi}}{\beta}\right)^{3/2}\sqrt{\frac32}\left(
1-\frac{2q^2}{3\beta^2}\right),$\\
$\mathcal{P}_3(q)=\left(\frac{2\sqrt{\pi}}{\beta}\right)^{3/2}
\sqrt{\frac{2}{15}}\left(\left(\frac{q}{\beta}\right)^4
-5\left(\frac{q}{\beta}\right)^2+\frac{15}{4}\right)$.
 one easily integrates over
$d^3q_1$ in (\ref{5.8}),(\ref{6.15}),(\ref{6.17}) with the result
\be J(\vep)=\bar Z(\vep) e^{-\frac{\vep^2}{\beta^2_2+2\beta^2_1} }
I_n (\vep), \label{6.19}\ee  and $I_n(\vep)$ is a polinomial in
$\frac{\vep^2}{\beta^2_1}$ with coefficients given in (A4.2),
(A4.3), with $\lambda=\frac{2\beta^2_1}{\beta^2_2+2\beta^2_1}$;
$\kappa^2=\frac{\beta_1^2\beta^2_2}{\beta^2_2+2\beta_1^2}$, and
$\bar Z(p)$ given in (\ref{5.9})\be J^{(1)}(\vep,\vek)=\frac{\bar
Z^{(\pi)}(\vep,\vek)}{\sqrt{2\omega_\pi(k)V_3}}
e^{-\frac{\vep^2}{\beta^2_2+2\beta^2_1}-\frac{\vek^2}{4\beta^2_2}}I_n(\vep).\label{6.20}\ee
Here $\bar Z^{(\pi)}(\vep,\vek)$ is given in (\ref{6.14}). Finally
the $2\pi$ matrix element is \be J^{(2)}(\vep,\vek_1,
\vek_2)=\frac{\bar Z^{(\pi\pi)}
 e^{-\frac{\vep^2}{\beta^2_2+2\beta^2_1}-\frac{\veK^2}{4\beta^2_2}}}{\sqrt{2\omega_\pi(k_1)V_32\omega_\pi(k_2)V_3}}
I_n(\vep),\label{6.21}\ee and $\bar Z^{(\pi\pi)}$ is given in
(\ref{6.16}).

Insertion of (\ref{6.19}), (\ref{6.20}), (\ref{6.21}) into
(\ref{6.5}) yields the form  ($i,k$ stand for polarizations of
initial and final $Q\bar Q $ states)
$$ w_{nn',ik}^{(\pi\pi)}(E)
=\frac{\gamma }{\sqrt{2\omega_\pi(k_1) V_3 2\omega_\pi(k_2)
V_3}}\frac{1}{f^2_\pi} \int\frac{d^3\vep}{(2\pi)^3}
e^{-\frac{\vep^2}{\beta^2_0}} I_n (\vep) I_{n'}(\vep)\times$$ 
 $$ \times \left\{ \left(
\frac{m^2_Q+\Omega^2}{2\Omega^2}\right)^2\frac{(\varphi_a\lambda_a)_1(\varphi_a\lambda_a)_2^*\delta_{ik}e^{-\frac{\vek^2_1+\vek^2_2}{4\beta^2_2}}}{E-E'(\vep)
-E(\vek_1 )}+(1\leftarrow 2)\right.$$\be\left.
-\left(\frac{m_Q}{2\Omega\omega_q}\right)^2{p_ip_k\vepi_1\vepi_2
e^{-\frac{\veK^2}{4\beta^2_2}}}\left[\frac{1}{E-E(\vep)-E(\vek_1,\vek_2)}+
\frac{1}{E-E(\vep)}\right]\right\}.\label{6.22}\ee

Here $\beta^{-2}_0=\left(
\frac{1}{2\beta^2_1+\beta^2_2}+\frac{1}{2\beta^{'2}_1+\beta^2_2}\right)$,
$E'(\vep)=\sqrt{\veP^2_1+M^2_s}+\sqrt{\veP^2_2+M^2_l},~~
\veE(\vep)=\sqrt{\veP^2_1+M^2_s}+\sqrt{\veP^2_2+M^2_s}$; $M_s=M_D$
or $M_B, M_l=M^*_D$ or $M^*_B$, $\veP_1\cong\vep-\frac{\veK}{2},~~
\veP_2\cong-\vep-\frac{\veK}{2}$.

To present our results in a convenient form, which can be compared
to experimental data, we shall use two standard variables on the
Dalitz plot, invariant mass $ M_{\pi\pi}$ of two pions, and cosine
of the angle $\theta$ of emitted pion $(\pi^+$ in $\pi^+\pi^-$)
$q^2\equiv M^2_{\pi\pi}=(\omega_1+\omega_2)^2 -\veK^2$, and pion
energies $\omega_1,\omega_2$ are expressed as \be
\omega_{1,2}=\frac{(M+M')\Delta M+ q^2}{4M}\mp
\frac{M+M'}{4M}\frac{\sqrt{q^2-4m^2_\pi}\sqrt{(\Delta
M)^2-q^2}}{q} \cos \theta.\label{89}\ee Moreover, the term
$\vek^2_1+\vek^2_1$ in the first exponential in (\ref{6.22}) is
written as  $$ \vek^2_1+\vek^2_1 = \frac{1}{8 M^2}\left\{\frac{}{}
((M+M') \Delta M+ q^2)^2+\right.$$ \be\left.+
(M+M')^2(q^2-4m^2_\pi) (\Delta M)^2-q^2)\frac{\cos ^2
\theta}{q^2}\right\} -2m^2_\pi \equiv\alpha +\gamma\cos^2 \theta.
\label{90}\ee

Now the phase space factor looks  like
$$
d\Gamma = \frac{d^3\vek_1d^3\vek_2}{(2\pi)^5} \frac{\delta
(\omega_1+\omega_2+
\sqrt{\veK^2+(M')^2}}{2\omega_12\omega_2}-M)=$$ \be =
\frac{(M+M')(M^2+M^{'2}-q^2)}{2(2M)^3 (2\pi)^3}\{ [(\Delta M)^2
-q^2][q^2-4m^2_\pi]\}^{1/2} dq d\cos \theta.\label{91}\ee The
second exponent in (\ref{6.22}) can be written as \be
-\frac{\veK^2}{4\beta^2_2}=-\frac{(\Delta
M)^2-q^2}{4\beta^2_2}\left(\frac{(M+M')^2-q^2}{4M^2}\right)\equiv
- \frac{(\Delta M)^2-q^2}{4\beta^2_2} \cdot
(1-\delta).\label{92}\ee

Now the basic elements in (\ref{6.22})(as will be shown later)
which define the form of the $\pi\pi$ spectrum can be written as
follows \be
\int\frac{d^3\vep}{(2\pi)^3}\frac{e^{-\frac{\vep^2}{\beta^2_0}}{I_n(p)
I_{n'}(p)p^{2k}}}{\frac{\vep^2}{2\tilde M}+E+\Delta M_{nn'}}\equiv
\frac{1}{\lan \frac{\vep^2}{2\tilde M}\ran +E +\Delta M_{nn'}}
\mathcal{F}^{(k)}_{nn'}\label{93}\ee \be \mathcal{F}^{(k)}_{nn'}
\equiv \int \frac{d^3\vep}{(2\pi)^3} e^{-\frac{\vep^2}{\beta^2_0}}
p^{2k}I_n(p) I_{n'}(p)=\beta_0^{2k}Q^{(k)}_{nn'}\label{94}\ee and
$Q^{(k)}_{nn'}$- dimensionless polinomials in ratios of $\beta_i$,
are  given  in Appendix 4.

Substituting (\ref{93}), (\ref{94}), (\ref{95}) in (\ref{6.22})
one finally obtains.
\be
 dw_{nn'} (q ,\cos \theta) \equiv d \Phi
 |\mathcal{M}|^2,\label{95}\ee
 where $d\Phi$ is
 \be d\Phi = \frac{1}{32\pi^3 N^2_c }\left( \frac{
 M_{br}}{f_\pi}\right)^4\frac{(M^2+M^{'2}-q^2)(M+M')}{4M^3}\sqrt{(\Delta
 M)^2-q^2}\sqrt{q^2-4m^2_\pi} d\sqrt{q^2} d\cos \theta
 \label{96}\ee
 with
 $$
 \mathcal{M}=\zeta\left\{\left(\frac{m^2_Q+\Omega^2}{2\Omega^2}\right)^2\left[
 \frac{1}{\frac{\lan \vep^2\ran}{2\tilde M^*}  + \omega_1 +\Delta
 M^*_{nn'}} + (1\leftrightarrow 2) \right]e^{-\frac{
 \alpha+\gamma cos^2\theta}{4\beta^2_2}}-\right.$$

 $$
 -\left(\frac{m_Q}{2\Omega \omega_2}\right)^2\left(
 \frac{\rho_{nn'}\beta^2_0}{3}\right) e^{-\frac{(\Delta
 M)^2-q^2}{4\beta^2_2}(1-\delta) }\left[\frac{1}{\frac{\lan
 \vep^2\ran}{2\tilde M} +\Delta M_{nn'}}+\right.$$

 \be\left.\left. + \frac{1}{\frac{\lan \vep^2\ran}{2\tilde M} +\omega_1
 +\omega_2 +\Delta M_{nn'}}\right]\right\}\label{97}\ee
Here we have defined  $\zeta=c_nc_{n'}p_{nn'}^{(0)}\left(
tt'\right)^{3/2}, $ $\Delta M^*_{nn'} = M_B+M_{B^*}-M(nS),$
$\Delta M_{nn'}= 2M_B-M(nS)$, $\delta=\frac{\Delta M}{M}
-\frac{(\Delta M)^2-q^2}{4M^2} \ll 1$ and the parameter, which
will be of importance, measuring the relative weight of $P$-wave
to $S$-wave intermediate states in the decay,

\be \rho_{nn'} =\frac{p^{(1)}_{nn'}}{p^{(0)}_{nn'}}.\label{98}\ee

The matrix element $\mathcal{M}$ is written for the case of the
intermediate (in most cases virtual) channel, e.g. for $ \Upsilon
(nS)\to \Upsilon(n'S)\pi\pi$ this channel is $BB^*$ for the first
square brackets, next channels are $B_sB_s^*$ and $B^*B^*$
respectively. In the next section we shall discuss general
properties  and numerical values of the obtained expressions.

\section{Results}

\subsection{The $\pi\pi$ spectrum}

It is convenient to rewrite $dw_{nn'}$ in (\ref{95}) as the
product of three characteristic factors: a combination of coupling
constants
$\frac{1}{\pi^4N_c^2}\left(\frac{M_{br}}{f_\pi}\right)^2$, phase
space factor \be d\Gamma_{ph} \equiv
\frac{(M^2+M^{12}-q^2)(M+M^1)}{4M^3} \sqrt{(\Delta
M)^2-q^2}\sqrt{q^2-4m^2_\pi} dq d\cos \theta\label{99}\ee and
matrix element squared, \be dw_{nn'}(q, \cos \theta)
=\frac{1}{32\pi^3N_c^2} \left(\frac{M_{br}}{f_\pi}\right)^4 d
\Gamma_{ph} |\mathcal{M}|^2.\label{100}\ee

We now evaluate all the factors in $ \mathcal{M}$,
Eq.(\ref{6.22}), for three typical transitions,
$$
A)~~~ \Upsilon(2S)\to \Upsilon(1S)\pi\pi$$

$$
B)~~~ \Upsilon(3S)\to \Upsilon(1S)\pi\pi$$

$$
C)~~~ \Upsilon(3S)\to \Upsilon(2S)\pi\pi.$$

The corresponding values of parameters  $\zeta$, $  \alpha,
\gamma, \rho_{nn'},\lan \vep\ran$ computed with the SHO wave
functions are given in Appendix 4.

Before going into the details of comparison of our results and
experiments, one can notice that the $\cos \theta$ dependence is
relatively weak and integrating over $\cos \theta $, and
neglecting weak dependence of $ \alpha, \gamma$ on $q$, one can
rewrite (\ref{97}) as follows:
$$ \mathcal{M}\equiv a(q)-b(q)\cong
a_{th}-b_{th}e^{\frac{q^2-4m^2_\pi}{4\beta^2_2}}=$$
$$=a_{th}-b_{th}
+b_{th}\left(1-e^{\frac{q^2-4m^2_\pi}{4\beta^2_2}}\right)=$$ \be
=\Delta_{th} - b_{th}
\left(\frac{q^2-4m^2_\pi}{4\beta^2_2}\right)\left(1+\frac{q^2-4m^2_\pi}{8\beta^2_2}+...\right)\label{101}\ee

Hence the character of the $\pi\pi$ spectrum for small $z\equiv
\frac{q^2-4m^2_\pi}{4\beta^2_2}$ is defined by the relation
between values of $\Delta_{th} =a_{th} -b_{th}$, and $b_{th}$
where $a_{th}=a(q=2m_\pi), b_{th} =b(q=2m_\pi)$.

It is convenient also to introduce dimensionless variable $x$,
changing in the interval [0,1] for all three transitions $A),B),
C)$ \be x=\frac{q^2-4m^2_\pi}{\mu^2},~~ \mu^2\equiv (\Delta
M)^2-4m^2_\pi ,\label{102}\ee and $\mu^2_A =0.234$ GeV$^2$,
$\mu^2_B=0.721$ GeV$^2$, $\mu_C^2=0.03$ GeV$^2$. In terms of $x$
the phase space is simple \be \frac{d\Gamma_{ph}}{dq}= \frac{(M^2+
M^{'2}-4m^2_\pi-x\mu^2)(M+M')}{4M^3}\mu^2 \sqrt{x(1-x)}d\cos
\theta\label{103}\ee which is roughly
$\frac{d\Gamma_{ph}}{dq}\approx \mu^2\sqrt{x(1-x)} d\cos \theta$.

Finally one can rewrite the  transition rate approximately and in
the SHO basis as \be
\frac{dw}{dq}=\frac{1}{32\pi^3N^2_c}\left(\frac{M_{br}}{f_\pi}\right)^4
\frac{\mu^6}{(4\beta^2_2)^2 }d\cos \theta \sqrt{x(1-x)}
b^2_{th}\left| \eta-x \left( 1
+\frac{\mu^2}{8\beta^2_2}x+...\right)^2\right|^2\label{104}\ee
with $\eta=\frac{4\beta^2_2}{\mu^2}\cdot
\frac{a_{th}-n_{th}}{b_{th}}$.One can see in (\ref{104}), that the
resulting $\pi\pi$ spectrum is defined by the last factor, which
provides three distinct types of behaviour, depending on the value
of $\eta$:

\begin{description}

   \item[] 1)~~  $|\eta|\ll1,~~
    \frac{dw}{dq\sqrt{x(1-x)}}\sim{x}^2$
    \item [] 2)~~ $\eta \sim \frac12 ,~~
    \frac{dw}{dq\sqrt{x(1-x)}}$ has a double peak and zero around
    $x\approx 0.5$.
    \item [] 3)~~ $\eta<-0.5~~
    \frac{dw}{dq} \sim \sqrt{x(1-x)}$, no significant structures
    in the spectrum.

\end{description}
As we shall see below, the behaviour { 1), 2), 3)} correspond to
the transitions $A),B), C)$ and this is supported by theoretical
estimates of $\eta$ in the SHO basis, and by experimental data.

We turn now to the evaluation of the coefficient $\eta$, using SHO
basis, with $a$ and $b$ given in (\ref{97}), and averaged over
$\cos\theta$, as it is done in $\pi\pi$ spectrum. \be
 a=\zeta\left(\frac{m^2_Q+\Omega^2}{2\Omega^2}\right)^2\left[
 \frac{1}{\frac{\lan \vep^2\ran}{2\tilde M^*}  + \omega_1 +\Delta
 M^*_{nn'}} + (1\leftrightarrow 2) \right]e^{-\frac{
 \alpha+\gamma cos^2\theta}{4\beta^2_2}}\label{105}\ee
\be
 b=\frac{\zeta}{3}\left(\frac{m_Q}{2\Omega \omega}\right)^2
 {\rho_{nn'}\beta^2_0}e^{-\frac{(\Delta
 M)^2-q^2}{4\beta^2_2}}\left[\frac{1}{\frac{\lan
 \vep^2\ran}{2\tilde M} +\Delta M_{nn'}}\right.
 \left.+ \frac{1}{\frac{\lan \vep^2\ran}{2\tilde M} +\omega_1
 +\omega_2 +\Delta M_{nn'}}\right],\label{106}\ee
and \be a_{th} \cong 2\zeta
\left(\frac{m^2_Q+\Omega^2}{2\Omega^2}\right)^2
e^{-\frac{\alpha+\gamma\cos^2\theta}{4\beta^2_2}}\xi^*_{nn'} ;~~
b_{th} = \frac{\zeta}{3}
\left(\frac{m_Q\beta_0}{2\Omega\omega}\right)^2 \rho_{nn'}
e^{-\frac{\mu^2}{4\beta^2_2}}(\xi_{nn'}+\tilde
\xi_{nn'})\label{107}\ee with \be \xi_{nn'}^*=\lan \frac{1}{\frac{
\vep^2}{2\tilde M} +\omega_i
  +\Delta M_{nn'}^*}\ran,\label{108}\ee  $$\xi_{nn'}=\lan \frac{1}{\frac{
\vep^2}{2\tilde M}   +\Delta M_{nn'}}\ran,$$
$$\tilde
\xi=\lan \frac{1}{\frac{ \vep^2}{2\tilde M}
  +\Delta M_{nn'}+\omega_1+\omega_2}\ran,$$
   exact values are defined in Appendix 4.

Finally the form of the spectrum as given in (\ref{104}) depends
on the value of $\eta(ns\to n's)$, which can be written as follows
\be \eta(nS\to n'S)=\left(\frac{2\lan e^{-\frac{ \alpha+ \gamma
\cos^2\theta
}{4\beta^2_2}}\ran_{\cos\theta}\xi^*_{nn'}u}{\frac13\rho_{nn'}
(\xi_{nn'} +\tilde
\xi_{nn'})e^{-\frac{\mu^2}{4\beta^2_2}}}-1\right)
\frac{4\beta^2_2}{\mu^2}\label{108a}\ee with
$u=\left(\frac{(m^2_Q+\Omega^2)~~\omega}{m_Q\Omega
\beta_0}\right)^2\approx \left(\frac{2\omega}{\beta_0}\right)^2,~~
\beta_0 (n,n')$ is given in Appendix 4.

The values of $\xi^*_{nn"}, \xi_{nn'}, \tilde \xi_{nn'}$ computed
with the SHO functions are given in Appendix 4 and the resulting
values of $\eta (nS\to n'S)$ are listed below in the Table 1.\\

\newpage

{\bf Table 1}\\ The values of the parameter $\eta$, computed with
SHO functions (upper line), with AZI restrictions (middle line)
and fitted to experiment using (115) (lower line).

\begin{center}

\begin{tabular}{|c|c|c|c|}\hline&&&\\
 $(nS\to n'S)$&  $2S\to 1S$& $3S\to 1S$& $3S\to 2S$\\\hline&&&\\
$\eta_{SHO}$&$\leq  0.45$&$\leq  0.27$&-3.66\\ \hline&&&\\
$\eta_{AZI}$ & 0.051&0.39&$-3.2 $\\&&&\\
\hline &&&\\ $\eta_{ (\exp)} (fit)$& 0& 0.52$\div 0.57$&$-2.7$\\
\hline

\end{tabular}

\end{center}

One can see, in (105)  that the magnitude of $\eta$ depends
strongly on the value of $\theta $, and we list in Table the
maximal values of $\eta_{SHO}$.

Comparison with experimental values can be done fitting $\eta $ to
three spectra in transitions  $A),B), C)$ which yields values
given in Table 1. One can see that the case $C)$ agrees quite
well, while in cases $A)$ and $B)$ the SHO values of $\eta$ in
table 1 are in the correct ballpark, however one needs a more
accurate calculation, since both  $\rho_{nn'}$ and $\eta_{nn'}$
depend strongly on the form of  wave functions, indeed the overlap
integrals in  matrix element enter there in the 4th power. The
obtained results and accuracy are enough only for our
semiquantitative analysis and need to be redone with realistic
wave functions. One way to obtain stable results is to  impose the
PCAC-Adler zero requirement on $ \mathcal{ M},$ which is done in
the next subsection.

\subsection{The PCAC improvements}

As was discussed in section 3, the cancellation  between the
tadpole double-pion amplitude (equivalent of our amplitude ``b'')
and the one-pion amplitude (equivalent of our ``a'') ensures, that
pion operators enter as $\partial_\mu\pi$ and hence satisfies the
Adler zero condition.
 In our discussion above a certain intermediate channels were
 chosen for a and b, different in general, and  for the
 resulting approximate
 amplitude $M$, Eq. (\ref{97}),it is not clear whether it does or  does not satisfy the Adler zero
 condition. To clarify the situation in this
 subsection we keep the form of amplitude (\ref{97}) however
 require the exact fulfillment of the Alder zero condition. To
 this end we rewrite the amplitude (\ref{97})  in terms of momenta $\vek_1,\vek_2$. One has thus the
 representation which will enable  us to obtain below the  ``Adler-zero-improved'' (AZI) form
 $$
 \mathcal{M}=const \left(\bar a~ e^{-\frac{\vek^2_1+\vek^2_2}{4\beta^2_2}}\left(
 \frac{1}{\frac{\lan \vep^2\ran}{2\tilde M^*} + \omega_1 +\Delta
 M^*_{nn'}}+ \frac{1}{\frac{\lan \vep^2\ran}{2\tilde M^*} + \omega_2+\Delta
 M^*_{nn'}}\right)-\right.$$

\be \left. \bar b ~e^{-\frac{(\vek_1+\vek_2)^2}{4\beta^2_2}}\left(
 \frac{1}{\frac{\lan \vep^2\ran}{2\tilde M} +\Delta
 M_{nn'}}+ \frac{1}{\frac{\lan \vep^2\ran}{2\tilde M} +\Delta
 M_{nn'}+ \omega_1+\omega_2}\right)-\right).\label{109}\ee

It is clear that when $\vek_1=\omega_1=0$, ($M$ is symmetric in
1,2), indeed $M$ vanishes whenever conditions are satisfied:

\begin{eqnarray}
&1)& \bar a (\vek_1=\omega_1=0,~~ k_2) = \bar b(k_1=\omega_1=0,
k_2)\\
\nonumber &2)&\frac{\lan \vep^2\ran}{2\tilde M^*} +\Delta
 M^*_{nn'}=\frac{\lan \vep^2\ran}{2\tilde M} +\Delta
 M_{nn'}\equiv \tau_{nn'}.\label{110}\end{eqnarray}
As a result one obtains for the $\pi\pi$ transition matrix element
a simple representation (the AZI form)
$$ \mathcal{M}_{nn'} =const
\left(e^{-\frac{\vek^2_1+\vek^2_2}{4\beta^2_2}}\left(
\frac{1}{\tau_{nn'}(\omega_1)+\omega_1}+\frac{1}{\tau_{nn'}(\omega_2)+\omega_2}\right)-\right.$$
\be\left. e^{-\frac{(\vek_1+\vek_2)^2}{4\beta^2_2}}\left(
\frac{1}{\tau_{nn'}(0)}+\frac{1}{\tau_{nn'}(\omega_1+\omega_2)+\omega_1+\omega_2}\right)\right).\label{111}\ee

It is clear, that (\ref{111}) satisfies the Adler zero condition
with any function $\tau_{nn'}(\omega)$. We have here two model
parameters: $\beta_2$, i.e. the radius of the $B,B^*$ (or $D,D^*$)
mesons, and $\tau_{nn'}$ -- the average transition energy. We fix
$\beta_2$ by the r.m.s. radius of $B$ meson (for $ \Upsilon$
transitions) as we did before, $\beta_2=0.5$ GeV, and $\tau_{nn'}$
calculate from the $\xi^*_{nn'}$ matrix element in Table 8.
Knowing average  values of $\omega_i$, given in Table 6, one
immediately obtains AZI coefficients $ \xi_{nn'}=\tau_{nn'}^{-1} $
and $ \tilde \xi _{nn'}=(\tau_{nn'} + \lan
\omega_1+\omega_2\ran_{nn'})^{-1}$.

Finally, expressing $\vek^2_i$ via $q^2$ and $cos \theta$ using
(\ref{90}), (\ref{92}) one obtains the AZI matrix element in the
form.

$$ \mathcal{M}_{nn'} =const
\left(e^{-\frac{\alpha+\gamma\cos^2\theta}{4\beta^2_2}}\left(
\frac{1}{\tau_{nn'}+\omega_1}+\frac{1}{\tau_{nn'}+\omega_2}\right)-\right.$$
\be\left. e^{-\frac{\mu^2(1-x)}{4\beta^2_2}}\left(
\frac{1}{\tau_{nn'}}+\frac{1}{\tau_{nn'}+\omega_1+\omega_2}\right)\right).\label{112}\ee

We now consider three typical cases of $\Upsilon(n)\to
\Upsilon(n')\pi\pi$. Here the values of $\xi^*\equiv
\frac{1}{\tau+\lan \omega\ran},~~ \xi=\frac{1}{\tau}$ and  $\tilde
\xi =\frac{1}{\tau+2\lan \omega\ran}$  are given in Table 8. They
are computed as in (93), and the Adler-zero-improved $\eta_{AZI}$
can be written as \be \eta^{(nn')}_{AZI} =
\frac{4\beta^2_2}{\mu^2} \left( \frac{2\xi^*_{nn'}}{\xi_{nn'}
+\tilde \xi_{nn'}} e^{\frac{\mu^2-\alpha-\gamma \cos^2
\theta}{4\beta_2^2}}-1\right).\label{114a}\ee

The resulting $\pi\pi$ spectrum has a simple approximate  form
\be\frac{dw}{dq} =const \sqrt{x(1-x)} |\eta(nn')
-x|^2\label{115a}\ee

The spectra $\frac{dw}{dq}$, corresponding  to (\ref{115a}) with
$\eta_{AZI}$ fitted to the data  are shown in Fig.6 together with
the experimental data of CLEO Collaboration \cite{42}. One can see
a reasonable argument for all three types of behaviour, $A), B),$
and $C)$with this simple parametrization. As it is seen in Table
1, the computed values of $\eta_{AZI}$ are very close to the
fitted
 ones.

\subsection{Total yield of $\pi\pi$}

In this subsection we derive the width of the dipion decays of
heavy quarkonia, $\Gamma_{\pi\pi}(nS\to n'S)$ using our general
equation (\ref{100}), (\ref{104})
$$\Gamma_{\pi\pi} (nS\to n'S)= \int dw_{nn'} (q,\cos \theta)=$$
\be =\frac{1}{32\pi^3N^2_c} \left(\frac{M_{br}}{f_\pi}\right)^4
\frac{\mu^7}{(4\beta^2_2)^2} b^2_{th} \int^1_0
\frac{dx\sqrt{x(1-x)}}{\sqrt{x+\frac{4m^2_\pi}{\mu^2}}}\left|
\eta-x \right|^2.\label{109v}\ee

We shall concentrate first on the $\Upsilon(2S)\to
\Upsilon(1S)\pi\pi$ transition, since here the SHO wave functions
might better imitate realistic ones, while for higher $nS$ states
the difference could be much larger.
\newpage



\begin{figure}
\vskip 0.5truecm \vspace{12pt}
 \hspace*{-0.2cm}
\includegraphics[width=20cm,height=22cm,keepaspectratio=true]{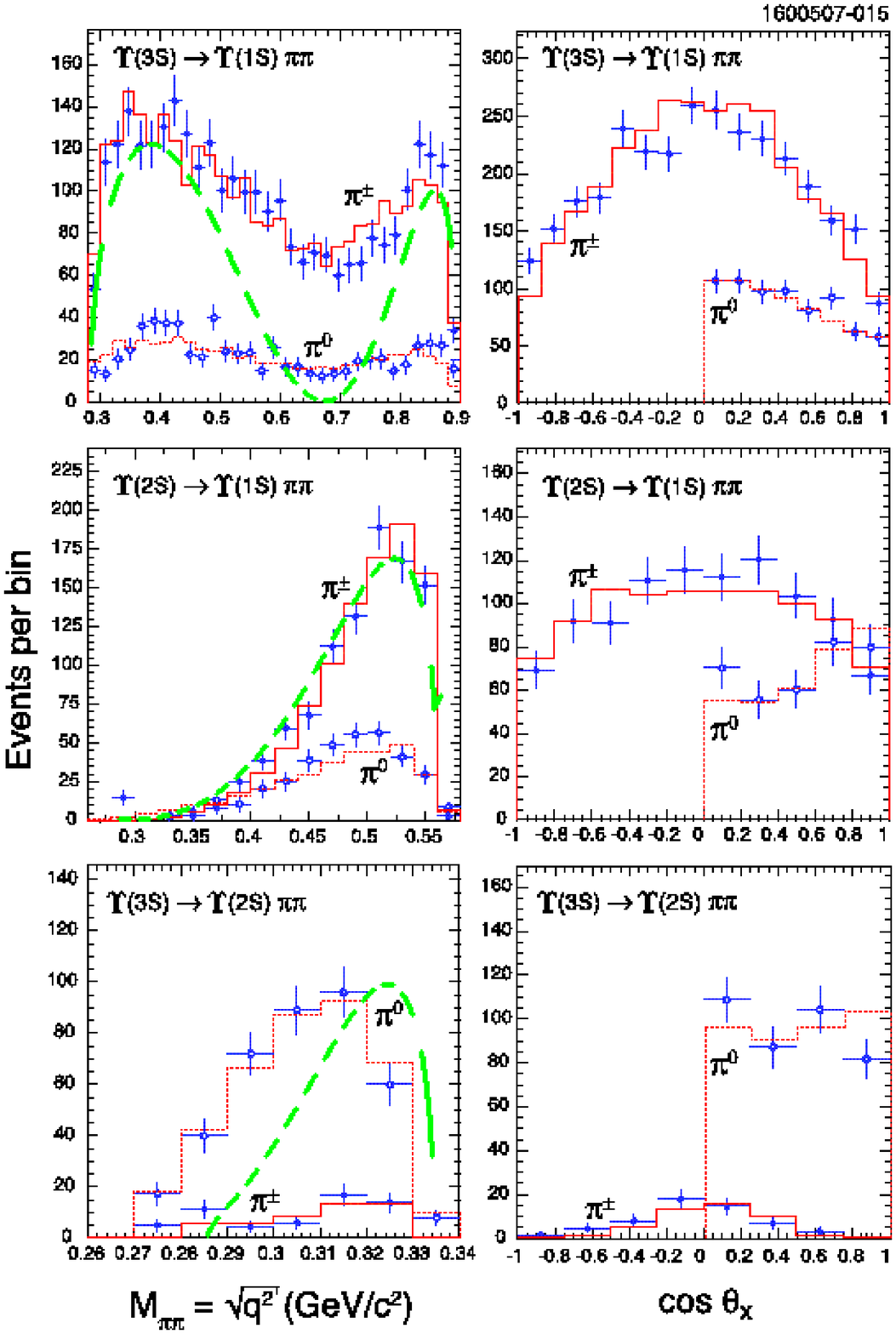}

\centerline{ \bf Fig.6}
\end{figure}

\newpage

Assuming first that  $\eta\ll 1$, we put it equal to zero (as it
follows from  the shape of spectrum in experiment). Then the phase
space integral with $x^2$ yields \be \int^1_0
\frac{x^{5/2}\sqrt{1-x}dx }{\sqrt{ x+\frac{4
m^4_\pi}{\mu^2}}}=0.116,~~{\rm for}~~\mu^2=0.234~{\rm~
GeV}^2,\Upsilon(2S)\to\Upsilon(1S).\label{110v}\ee This yields \be
\Gamma_{\pi\pi}(2S\to1S)=\frac{1}{32\pi^3N^2_c}\left(\frac{M_{br}}{f_\pi}\right)^4\frac{\mu^7}{(4\beta^2_2)^2}
\cdot 0.116 b^2_{th}.\label{110vv} \ee  with $b_{th} (2S\to 1S)=
0.92\zeta$ and $\zeta $ is given in Table in Appendix 4,
$\zeta^2(2S\to 1S)=0.194$. As a result one obtains \be
\Gamma_{\pi\pi} (2S\to 1S) =\left(\frac{M_{br}}{f_\pi}\right)^4
13\cdot 10^{-9} {~~\rm
GeV}=0.013\left(\frac{M_{br}}{f_\pi}\right)^4{~\rm
keV}.\label{111v}\ee

Taking $M_{br}\cong 1$  GeV  as follows from $\Gamma(\psi(3770)\to
D\bar D)$, see section 5, one obtains $\Gamma_{\pi\pi}\approx
171~$ keV, while  experimental value  is \cite{62} $\Gamma_{\exp}
(\Upsilon(2S)\to \Upsilon (1S)\pi^+\pi^-)=(8.1\pm 1.2)$ keV. This
distinction  is not surprising, taking into account the fourth
power of $\left(\frac{M_{br}}{f_\pi}\right)$ and also the fourth
power of overlap integrals between $\Upsilon(nS)$ and $B\bar B$
wave function.

In particular, if one fits the value of $\beta_1$ for
$\Upsilon(2S)$ to reproduce the position of zero in the realistic
$x$-space wave function (\cite{18,61},  and private communication
by the authors), then $\beta_1$ will increase 1.6 times as
compared to that fitted to the r.m.as. radius of $\Upsilon (2S)$,
and the resulting value of $\zeta^2$ and hence of
$\Gamma_{\pi\pi}(2S\to1S)$ decreases twenty times (!).
Consequently, $\Gamma_{\pi\pi}^{(zero~fitted)} (2S\to 1S)\approx
8.5$ keV, which is comparable to the experimental result.

To check, whether the mass dependence of our  formulas  is
correct, we now turn to the  case $\psi(2S)\to J/\psi\pi\pi$.
Calculating again with the SHO wave functions, one obtains
(fitting the r.m.s. radii 0.4 fm for $J/\psi$, 0.8 fm for
$\psi(2S)$, and 0.58 fm for $D$) $\beta'_1=0.94$ GeV,
$\beta_1\equiv \beta_1(\psi(2S))=0.47$ GeV, $\beta_2=0.426$ GeV.
As a consequence (see Appendix 4 for details) one has
$\rho^\psi_{21}= 2.66$, and  $\zeta^\psi=-0.454$, and
$b_{th}=-0.11$ GeV$^{-1}$ and the result is $$
\Gamma^\psi_{\pi\pi}=3.8 \left(\frac{M_{br}}{f_\pi}\right)^4
10^{-9}~{\rm GeV}.$$ Taking again $M_{br}\cong 1$ GeV, as fitted
to the case $\psi(3770)\to D\bar D$, see  section 5, one finally
gets $\Gamma_{\pi\pi}^\psi=51$ keV. This should be compared with
PDG data, $\Gamma_{\exp} =158 $ keV. One can see that our
theoretical value is within factor of 3 from the experimental
result, which is reasonable, taking into account the crudeness of
the SHO wave functions. Again, since the SHO result is
proportional to $\frac{1}{\beta^4_2}\sim (r_D)^4$, where $r_D$  is
the radius of $D$ meson, changing $\beta_2$ by 40\% one obtains
the 4 times increase  of $\Gamma^\psi_{\pi\pi}$, making it
comparable to the experimental value.

We do not compare total yields for higher $\Upsilon(nS)$, with
$n\geq 3$, since there the total SHO  yield becomes too large due
to specific properties of SHO vawefunctions, where  $\lan
p^n\ran_{SHO}$ grow almost factorially with $n$, which is
unrealistic.

The resulting dipionic spectra of $\Upsilon(nS)\to \Upsilon
(n'S)\pi\pi$ with $n=2,3$ and $n'=n-1,$ $n-2$, are given in Fig. 6
in comparison with the CLEO results from \cite{42}.

One can notice that the one-parameter fit of the form (\ref{111})
is working reasonably well, yielding values of $\eta_{AZI}\approx
\eta_{\exp}$  (fit) in Table 1. These values of $\eta_{\exp}$ are
in the same ballpark as the theoretical values $\eta^{SHO}$ given
in the same Table 1.

At this point it is important to stress that in the fits  and
estimates it was the value of $\eta$ averaged over angle $\theta$
that was used, as in Eq.(\ref{108a}).  However, $\eta$ has
 the $\theta$ dependence visualized in Eq.(\ref{105}), and it can
 be written as
 \be\frac{\mu^2}{4\beta^2_2} \eta=\lan \frac{a}{b}\ran_\theta
 e^{-\frac{\gamma}{4\beta^2_2}(\cos^2\theta-\cos^2\bar
 \theta)}-1\approx \lan \eta\ran_\theta+(\cos^2\bar
 \theta-\cos^2\theta)\frac{
 \gamma}{4\beta^2_2}.\label{120}\ee
 Therefore 1) a strong $\theta$ dependence appears when
 $\frac{\gamma}{4\beta^2_2}$ is large, as in the $3S\to 1S$
 case, when $\frac{\gamma}{4\beta^2_2}=0.4$, and the
 enhancement at $\theta\approx 0$ is explained seen in the CLEO
 data \cite{42}. In $2S\to 1S$, $\frac{\gamma}{4\beta^2_2}\approx
 0.16$ and enhancement is more shallow.

 2) The dip in the $\pi\pi$ spectrum in ($3S\to 1S)$ case is at
 $x=\eta(\theta)$ and depends on $\theta$. integrating the
 spectrum over $d\theta$ one is summing over different dip
 positions, which results in a partial filling up the dip, in
 agreement with experiment, see Fig. 6.

 \section{Discussion}

 Our strategy in this paper  is drastically different from the
 generally accepted ``multipole expansion -- Adler zero''
 approach. The latter is essentially exploiting the idea of the
 small-size source of
gluon fields. In particular the approach of the ITEP group
\cite{17,31,36} is aesthetically appealing, starting with the
local condensate of gluonic field, turning it into the
energy-momentum tensor of gluons and finally into that of pions,
predicting in this way a peak at  high $\pi\pi$ mass and a correct
ratio of $\eta/\pi \pi$ yield in  $(2S\to 1S)$.

Our argument is that this might well be true for small size
systems, such as toponium, however becomes  inapplicable for
systems of size $R$ larger than the vacuum correlation length
$\lambda\approx 0.2$ fm \cite{19}. Actually all decaying systems
in question have r.m.s. radii not smaller than 0.4 fm, when
nonlocality of gluon condensates is vitally important and brings
confinement (see \cite{43} for discussion). It was checked in
\cite{63} that  the local gluonic condensate strongly
overestimates  binding energy of all states with $R\geq 0.4$ fm,
while nonlocal ones (confinement) yields  correct masses of
$\Upsilon(nS)$ with $n\geq 1$.

Therefore our approach is an attempt in applying  nonperturbative
QCD methods to the large size systems, using derived   in this way
quark-pion  Lagrangian \cite{46}. It contains the only parameter-
$M_{br}$, which we fix by independent input -- the width of
$\psi(3770)\to D\bar D$, yielding $M_{br}\approx 2\omega\approx 1$
GeV.

The $\pi\pi$ production in our approach is due to the presence of
light $q\bar q$ pair in the body of heavy quarkonium, which should
be $O(1/N_c)$ in amplitude. Each pion is emitted with
$\frac{1}{f_\pi}$ amplitude, which yields $\pi\pi$ production
amplitude
$O\left(\frac{1}{N_cf^2_\pi}\right)=O\left(\frac{1}{N_c^2}\right)$.
One can see in (\ref{109}), that indeed
$\Gamma_{\pi\pi}=O\left(\frac{1}{N^2_cf^4_\pi}\right)=O\left(\frac{1}{N^4_c}\right)$.

Furthermore, when the $q\bar q$ pair is formed inside
$\Upsilon(nS)$ the latter lives part of time in $B\bar B$ or
$B\bar B^*$ (or else $B^*\bar B^*$)states.  The quark-pion
Lagrangian (\ref{38b}) contains vertices with emission of any
number of pions (or kaons and etc.),  and  essentially there
appear two types of amplitudes: (a) with one-pion emission at each
vertex, see Fig. 4, (b) with two-pion emission at one vertex  and
zero-pion emission at another, see Fig. 5 .

It was shown in the  paper, that due to chiral properties of
quark-pion Lagrangian (essentially equivalent to the  Adler zero
requirement) these two amplitude enter with different signs,
$M=a-b$.

Finally, a and b, when computed with any reasonable (decreasing
with argument) functions are \underline{decreasing} functions of
different pionic variables, namely, $a=a(\vek^2_1,\vek^2_2),~~
b=b((\vek_1+\vek_2)^2)$, in case of SHO functions, $a^{SHO}\sim
\exp \left(-\frac{\vek^2_1+\vek^2_2}{4\beta^2_2}\right),~~
b^{SHO}\sim
\exp\left(-\frac{(\vek_1+\vek_2)^2}{4\beta^2_2}\right)$.

This fact implies completely different dependence in terms of
$q^2=(k_1+k_2)^2$ and $\cos \theta$, indeed, b does not  depend on
$\theta$ and grows with $q^2,~~ b^{SHO}\sim
\exp\frac{q^2}{4\beta^2_2}$, while  $a^{SHO}\sim \exp
\left(-\frac{\alpha(q)+\gamma(q)\cos^2\theta}{4\beta^2_2}\right)$.
The same type of behavior is expected for realistic
wave-functions, $a\sim f_a(\alpha (q),
\gamma(q)\cos^2\theta),b\sim f_b(q^2),$ with $f_a$ decreasing and
$f_b$ increasing functions of their arguments.  The last feature
is the cancellation of $a_{th} =a(q^2=4m^2_\pi)$, and $b_{th}$--
the complete  cancellation in the case $A)$ $(2S\to 1S)$, a
partial cancellation in case $B)~~ (3S\to 1S)$, and no
cancellation at all in case $C)~~(3S\to 2S)$. This feature is
qualitatively supported by the SHO calculations (see Table 1), and
needs to be confirmed by the realistic calculation.

As a result one  obtains a simple  one-parameter representation of
the $\pi\pi$ spectrum which works well in case of
$\Upsilon(2S),\Upsilon(3S)$ and $\psi(2S)$ as shown in this paper.
Moreover, one finds explanation not only for the form of the
spectrum, as shown in Fig.6, but also qualitatively for the $\cos
\theta$ dependence.

The extrapolation of the method to the cases of $\Upsilon(4S)$
decays \cite{25,26} and to the $\Upsilon(5S)$ transitions, found
recently by Belle \cite{64}, is straightforward. One can easily
see, that  the corresponding $\pi\pi$ spectra are easily described
by  $\eta$ parametrization, however the SHO calculation of spectra
is hardly applicable for such high $n,n'$, and  one needs
realistic wave functions and possibly more channels of $B, \bar B$
type should be included. This work is planned for the future.

\section{Summary and outlook}

\begin{description}
    \item[] 1. We have constructed the amplitudes, $\pi\pi$ spectra
    and the total width $\Gamma_{\pi\pi}$ for all dipionic
    transitions  $\Upsilon(nS)\to \Upsilon(n'S)\pi\pi$ with
    $n=2,3$, and  $n'=1,2$. The only parameter of our approach,
    $M_{br}$ in the effective  chiral Lagrangian, $\int\bar
    \psi(x)M_{br}\exp (i\gamma_5\hat \phi) \psi(x) d^4x$ is fixed
    using the decay $\psi(3770)\to D\bar D$.
    \item[] 2. It is shown, that dipion spectra have the form
    $d\Gamma_{\pi\pi}=$ phase space times  $|\eta-x|^2$,  where $\eta \sim \frac{a_{th}-
    b_{th}}{b_{th}}$ and $a_{th}$  refers to the  threshold  amplitude with sequential
    decay of the type $\Upsilon(nS)\to \bar B B^*\pi\to
    \Upsilon(n'S)\pi\pi$, and $b_{th}$ -- the amplitude for the
    two-pion-vertex emission: $\Upsilon(nS)\to B\bar B\pi\pi\to
    \Upsilon(n'S)$ plus $\Upsilon(nS)\to B\bar B\to
    \Upsilon(n'S)\pi\pi$.
    \item[] 3. It is shown, that $\pi\pi$ spectra appear of three
    kinds: $A)$ when  $|\eta|\ll 1$ the
    spectrum behaves roughly as  $(q^2-4m^2_\pi)^2$, as in ($2S\to
    1S$) transition in $\psi(2S)$ and $\Upsilon(2S),~~B)$ when
    $\eta\sim \frac12$ there appears a dip in the middle of spectrum,
    as in the case $\Upsilon (3S\to 1S), \Upsilon(4S\to 2S)$,
    $C)$ when $\eta<-0.5$, there is a moderate enhancement at
    large $q^2$ due to amplitude b, as in the  $\Upsilon(3S\to
    2S)$. The case $\Upsilon(4S\to1S)$ seemingly belongs to case
    $B)$ with $\eta_{\exp}(4S\to 1S)\approx 0.30$ and $\eta_{\exp}(4S\to 2S)\approx 0.61$.
    \item[]4. We have checked the method, computing absolute
    normalization of $\psi(2S\to 1S)\pi\pi$ and $\Upsilon  (2S\to
    1S)\pi\pi$ using SHO functions and found qualitative
    agreement.  We also computed $\eta(\cos \theta)$  with the
    same SHO functions with parameters fitted to the known r.m.s.
    radii of all states. Remarkably, in spite of crudeness of
   this approximation, we  have  found a reasonable agreement with
   experimental $\pi\pi$ spectra, especially  when the Adler zero condition is imposed (see $\eta_{AZI}$ in Table 1 compared to
   $\eta_{exp}$(fit)
   ).

    \item[] 5.~ We have argued, that multipole expansion is not
    suited for all $nS$ states with $n>1$, since their size is
    larger than correlation length of gluonic  vacuum,
    $\lambda=0.2$ fm.
    \item[] 6.~~ We have casted  some doubt on the use of PCAC
    vanishing of amplitude as a sole  source of damping of $\pi\pi$ spectrum , giving example of amplitude with Adler
    zero, but actually not changing across available phase  space.

    Moreover, the $(q^2-4m^2_\pi)^2$ behavior in $2S\to 1S$ is not
    due to PCAC. Still, relative sign of a and b is necessary for the Adler zero condition
    and is vitally important for description of $\pi\pi $ spectra.
    \item[] 7. ~It is argued that $\pi\pi$ FSI is not affecting
    much the form of spectrum, in $\Upsilon (nS), n<4$  since both major features: the
    $(q^2-4m^2_\pi)^2$ damping at threshold for $(2S\to 1S)$ and
    the dip in $(3S\to 1S)$ are seemingly not connected to FSI.
    However, FSI might be important for the part of spectrum
    around $q\sim 1$ GeV.
    \item[] 8.~ ~In developing the formalism, we have suggested the
    method of constructing the relativistic amplitudes of
    multipoint and multihadron types, the latter expressed via
    eigenfunctions of relativistic Hamiltonian.

    \item[] 9.~  The same method is easily  generalized for the
    case of scattering amplitude of pion on heavy quarkonia, and
    preliminary study reveals a strong interaction, which may be
    of importance for the explanation of the states like
    $Z(4430)$, recently discovered by Belle \cite{65}.

 \item[]10.~ ~For $n\geq5$ the $\Upsilon (5S\to n'S)$ transitions
 proceed via the open $B\bar B$ channels, and our Eq. (\ref{6.22})
 predicts a strong absorptive part in the amplitude $ \mathcal{M}$
 (97) which is proportional to the $4\pi p_BM_B$ and produces an
 enhancement factor $\sim 10^3$ in the total $\pi\pi$ width, in
 experiment
 $\Gamma^{\exp}_{\pi\pi}(5S\to 1S)\sim 1 $ MeV \cite{64}, whereas  $\Gamma^{\exp}_{\pi\pi}(3S\to 1S)\sim 1 $
 keV \cite{42}. This fact  gives an additional support to the mechanism
 proposed in the paper.
\item[]11.~ ~ As it is clear from (\ref{111v}) the main dependence
of total width on the mass of emitting Nambu-Goldstone mesons
comes from the $\mu^{7}$ dependence. Hence for the $\Upsilon
(5S\to 1S)$ transitions the ratio $\frac{\Gamma_{K^+K^-}(5S\to
1S)}{\Gamma_{\pi\pi} (5S\to 1S)} \sim \left(
\frac{\mu_K}{\mu_\pi}\right)^{7}\approx 0.104$  $
(\mu_i=\sqrt{(\Delta M^2)- 4 m^2_i})$, which roughly agrees with
recent experimental data \cite{64}.

\end{description}

   \section{Acknowledgements}

   The author is indebted to A.M.Badalian and B.L.G.Bakker for
   providing data, and to A.M.Badalian for many useful
   discussions. The author is grateful to E.V.Komarov for help in
   preparing the Fig.6, to K.G.Boreskov, A.B.Kaidalov, Yu.S.Kalashnikova and V.I.Zakharov  for
   discussions. Useful discussions with M.V.Danilov and P.N.Pakhlov
   were helpful for the author.

   The financial support of RFFI grant 06-02-17012 and the grant for
   scientific schools NSh-843.2006.2 is gratefully acknowledged.

\newpage

\vspace{2cm}

{\bf Appendix 1}\\

{\bf Factorization of the Dirac bispinorial structure of the string breaking Green's function $G_{br}$ }\\

 \setcounter{equation}{0} \def\theequation{A1.\arabic{equation}}

Writing in (\ref{40z}) $J_{Q\bar Q} \equiv
\frac{tr}{\sqrt{N_c}}\bar \psi_Q (\veu) \Gamma_Q\Phi(\veu,\vev)
\psi_Q(\vev)$ and  \be J_{Q\bar Q}^+(\veu',\vev') \equiv
\frac{tr}{\sqrt{N_c}}\bar \psi_Q (\veu') \bar
\Gamma_Q\Phi(\veu',\vev') \psi_Q(\vev'),\label{A.1}\ee and for the
light quark vertices \be J_{q\bar q} (x) \equiv \bar \psi_q(x)
\Gamma_x\psi_q(x), ~~ J^+_{q\bar q}(y) \equiv \bar \psi_q (y) \bar
\Gamma_y\psi_q(y)\label{A.2}\ee one has the following structure
for a given choice of two intermediate heavy-light bosons $( \bar
\psi_Q(w) \Gamma_{Qq} \psi_q(w)), ~~ (\bar \psi_q(w')
\Gamma_{qQ}\Psi_Q(w')$, as depicted in the left part of  Fig.7
(the right part has a similar structure).

\vspace{1cm}

\unitlength 1mm 
\hspace{1cm}
 \linethickness{0.4pt}
\ifx\plotpoint\undefined\newsavebox{\plotpoint}\fi 
\begin{picture}(78.75,51.75)(0,0)
\put(17.25,16.5){\framebox(55.5,30.25)[cc]{}}
\put(17,32.25){\circle*{2.83}} \put(72.5,32.75){\circle*{2.5}}
\put(44.5,47.25){\circle*{3.16}} \put(45.75,16.5){\circle*{.5}}
\put(45.5,16.5){\circle*{3.04}}
\multiput(17.75,33)(.0328947,.0855263){38}{\line(0,1){.0855263}}
\multiput(19,36.25)(.0333333,.05){45}{\line(0,1){.05}}
\multiput(20.5,38.5)(.0457317,.0335366){82}{\line(1,0){.0457317}}
\multiput(24.25,41.25)(.0583333,.0333333){60}{\line(1,0){.0583333}}
\multiput(27.75,43.25)(.1447368,.0328947){38}{\line(1,0){.1447368}}
\multiput(33.25,44.5)(.158333,.033333){30}{\line(1,0){.158333}}
\multiput(38,45.5)(.1201923,.0336538){52}{\line(1,0){.1201923}}
\multiput(17.25,32.75)(.0335821,-.0895522){67}{\line(0,-1){.0895522}}
\put(19.5,26.75){\line(0,1){0}}
\multiput(19.5,26.75)(.0335821,-.0447761){67}{\line(0,-1){.0447761}}
\multiput(21.75,23.75)(.0666667,-.0333333){60}{\line(1,0){.0666667}}
\multiput(25.75,21.75)(.1,-.0333333){45}{\line(1,0){.1}}
\multiput(30.25,20.25)(.1381579,-.0328947){38}{\line(1,0){.1381579}}
\multiput(35.5,19)(.35,-.033333){15}{\line(1,0){.35}}
\multiput(40.75,18.5)(.1055556,-.0333333){45}{\line(1,0){.1055556}}
\put(45.25,17){\line(4,1){5}}
\multiput(50,18.75)(-.03125,-.0625){8}{\line(0,-1){.0625}}
\multiput(49.75,18.25)(.2105263,.0328947){38}{\line(1,0){.2105263}}
\put(57.75,19.5){\line(3,1){5.25}}
\multiput(63,21.25)(.0457317,.0335366){82}{\line(1,0){.0457317}}
\multiput(66.75,24)(.0335366,.0396341){82}{\line(0,1){.0396341}}
\multiput(69.5,27.25)(.0335821,.0858209){67}{\line(0,1){.0858209}}
\put(44,48){\line(1,0){.25}}
\multiput(44.25,48)(.03125,-.03125){8}{\line(0,-1){.03125}}
\multiput(44.25,47.25)(.15,-.0333333){75}{\line(1,0){.15}}
\multiput(55.5,44.75)(.0777778,-.0333333){45}{\line(1,0){.0777778}}
\put(59,43.25){\line(5,-3){5}}
\multiput(64,40.25)(.0433333,-.0333333){75}{\line(1,0){.0433333}}
\multiput(67.25,37.75)(.0365854,-.0335366){82}{\line(1,0){.0365854}}
\multiput(70.25,35)(.0333333,-.05){45}{\line(0,-1){.05}}
\put(13.5,47.75){u} \put(13.75,15.5){v} \put(22.25,32.25){x}
\put(44,51.75){w} \put(78.75,33.25){y} \put(77.75,47.75){u'}
\put(78.5,15.25){v'} \put(44.75,21.25){w'} \put(43.5,8.5){Fig 7}
\end{picture}

\be G_{Q\bar Q, q\bar q}\equiv tr_L (\Gamma_Q S_Q(u, w)
\Gamma_{Qq} S_q(w,x) \Gamma_q S_q(x, w')\Gamma_{qQ}
S_Q(w',v)).\label{A.3}\ee Here $tr_L$ denotes the trace over Dirac
bispinor indices. In the FFSR one writes the quark Green's
function as (in Euclidean notations), $ k=q,Q$ $$S_k(x,y)=
(m_k-\hat D) (m^2_k-\hat D^2)^{-1}=$$ \be= (m_k-\hat D)
\int^\infty_0 ds (Dz)_{xy} e^{-K} \Phi(x,y) \exp (g\int^s_0 d\tau
\sigma_{\mu\nu} F_{\mu\nu})\label{A.4}\ee

The last factor in (\ref{A.4}) takes into account spin-dependent
interaction and can be disregarded in the first approximation.
Then all Dirac matrix structure of (\ref{A.4}) is furnished by the
prefactor $(m_k-\hat D)$, which can be rewritten in terms of
momentum as shown in \cite{57}, namely for a simple
quark-antiquark loop one has \be Y_\Gamma=\frac14 tr_L(m_1-\hat
D_1)\Gamma(m_2-\hat D_2); ~~\Gamma=\frac14 tr_L((m_1-i\hat
p_1)\Gamma(m_2+i\hat p_2)\Gamma.\label{A.5}\ee Here $\hat
p_i=p_\mu^{(i)}\gamma_\mu, ~~i=1,2,$ and in the c.m. system
$p_i^{(1)}=-p_i^{(2)}=p_i,~~ p_4^{(1)}=i\omega_1,
~~p_4^{(2)}=i\omega_2$, so that\be Y_\Gamma=\frac14 tr_L
(\Gamma(m_1+\omega_1\gamma_4-ip_k\gamma_k)\Gamma
(m_2-\omega_2\gamma_4-ip_i\gamma_i)).\label{A.6}\ee

Note, that the index 1  is  referred to a quark, while index 2 to
an antiquark. In (\ref{A.3}) one has heavy quark $Q$ with momentum
$p_\mu^{(Q)}=(i\omega_Q, p_i^{(Q)}),$
 heavy antiquark $\bar Q$
with momentum $p_\mu^{(\bar Q)}=(i\bar\omega_Q, \bar p_i^{(Q)}),$
light quark $q$ with momentum $q_\mu=(i\omega_q, q_i)$, and light
quark $\bar q$ with momentum $\bar q_\mu=(i\bar\omega_q, \bar
q_i)$. Writing $S_k(x,y)= (m_k-\hat D) G_k(x,y)$ and taking all
$G_k$ out of the sign of  $tr_L$, since they are proportional to
the unit matrix in Dirac indices, one has  $G_{Q\bar Qq\bar
q}\equiv Z\Pi_kG_k$, where
$$ Z=tr_L[(\Gamma_Q(m_Q+\omega_Q\gamma_4-ip_i^{(Q)}\gamma_i)\Gamma_{Qq}(m_q-\bar\omega_q\gamma_4+iq_k\gamma_k)$$
\be\Gamma_q(m_q+\omega_q\gamma_4-iq_l\gamma_l)\Gamma_{Qq}(m_{\bar
Q}-\bar\omega_{\bar Q}\gamma_4+i\bar p^{(Q)}_n\gamma_n))]
.\label{A.7}\ee

Matrices $\Gamma_i, ~~i=Q,Qq,q,qQ$, define the quantum numbers of
participating particles, while $\Gamma_q$ can be only of two
types: $\Gamma_q=1$ for a $q\bar q$ vertex without NG emission or
emission of even numbers of NG meson
 and $\Gamma_q=i\gamma_5\frac{\varphi_a\lambda_a}{f_\pi}$ for one NG meson emission, and $-\frac{\varphi_a
 \lambda_a\cdot \varphi_b\lambda_b}{2f^2_\pi}$ for two NG meson emission.

 To normalize the factor $Z$ properly, one should take into account, that the path integral for the scalar part
 of the quark Green's function $G_k(x,y),~~ k=q,Q$ can be rewritten as
 follows \cite{57}

\be G_k(x,y)=\int^\infty_0 ds (D^4z)_{xy}e^{-K}
\Phi(x,y)=\int\frac{(D^3z)_{\vex\vey}}{2\omega_k} D\omega e^{-K}
\Phi(x,y),\label{A.8}\ee where $\omega_k$ is an averaged energy of
the quark, $\omega_k=\lan\sqrt{m^2_k+\vep^2_k}\ran$,and for the
white quark-antiquark Green's function one has
$$
\frac{1}{N_c}\lan tr_cG_k(y,x)\Phi(x,\bar x) G_{\bar k}(\bar
x,\bar y)\Phi(\bar y,y)\ran=\frac{1}{4\omega_k\omega_{\bar k}}
\lan x,\bar x|e^{-HT}|y,\bar y\ran=$$ \be
=\frac{1}{4\omega_k\omega_{\bar k}}\sum_n\Psi_n (x,\bar
x)\Psi^+_n(y,\bar y)e^{-E_nT}.\label{A.9}\ee

Therefore it is convenient to introduce the projection factors
$\Lambda^\pm_k$ where subscripts $(+)$ and $(-)$ stand for quarks
and antiquarks respectively \be
\Lambda^\pm_k=\frac{m_k\pm\omega_k\gamma_4\mp ip^{(k)}_i
\gamma_i}{2\omega_k},~~ k=q,Q,\label{A.10}\ee and the normalized
factor $\bar Z$ looks like \be \bar
Z=tr_L(\Gamma_Q\Lambda^+_Q\Gamma_{Q\bar q}
\Lambda^-_q\Gamma_q\Lambda^+_q\Gamma_{q\bar
Q}\Lambda_Q^-)\label{A.11}\ee Using (\ref{A.11}) and
(\ref{a1'.23}) arrives at Eq.
  (\ref{43})

As a next step one separates c.m.momenta and introduces in the
$Q\bar Q$ system Green's function with zero c.m. momentum and in
the energy representation $G_{Q\bar Q}^{\veP=0} (\ver,\ver', E)$
as in (\ref{41}) and the same in the $Q\bar Q$ system. Note that
dimension of $G_{Q\bar Q}$ is $[G_{Q\bar Q}^{\veP=0} (\ver,\ver',
E)]=m^2$.

Separation of c.m. and relative coordinates goes as follows.
Consider $Q\bar Q q\bar q$ system represented as a system of two
bound states $ Q \bar q$ and $\bar Q q$ with c.m. coordinate
$\veR_1$ and $\veR_2$ and c.m. momenta $\veP_1$ and $\veP_2$. One
can find total c.m. and relative momentum of the $Q\bar Q q\bar q$
system as $\veP=\veP_1+\veP_2,~~\vep=\frac12 (\veP_1-\veP_2)$, and
c.m. and relative coordinates of equal mass $Q\bar q$ and $\bar
Qq$ mesons as $\veR=\frac{\veR_1+\veR_2}{2},~~\ver=\veR_1-\veR_2$,
so that one has $\veP_1\veR_1+\veP_2\veR_2=\veP\veR+\vep\ver$.

On the other hand, c.m. coordinates are found via the average
energies of quarks, so that in the initial and final states of the
amplitude depicted in Fig.7  one has \be \veR_1 = \frac{\omega_1
\vex+\Omega_1 \veu}{\omega_1+\Omega_1},~~~\veR_2 = \frac{\omega_2
\vex+\Omega_2 \vev}{\omega_2+\Omega_2}\label{A.12}\ee

\be \veR'_1 = \frac{\omega_1 \vey+\Omega_1
\veu'}{\omega_1+\Omega_1},~~~\veR'_2 = \frac{\omega_2
\vey+\Omega_2 \vev'}{\omega_2+\Omega_2}.\label{A.13}\ee

For the same masses in $Q\bar q$ and $\bar Q q$,
$\omega_1=\omega_2, \Omega_1=\Omega_2$ one finds \be
\veR=\frac{\omega \vex}{\omega+
\Omega}+\frac{\Omega}{\omega+\Omega} \frac{\veu+\vev}{2}, ~~ \ver
=\frac{\Omega}{\omega+\Omega}(\veu-\vev)\equiv
c(\veu-\vev)\label{A.14}\ee

As a result the total integration element is $$ d^3\vex d^3 \veu
d^3\vev=\left(\frac{\Omega+\omega}{\Omega}\right)^3 d^3\veR
d^3\ver d(\vex-\veu)=$$ \be d^3\veR
d^3(\vev-\veu)d^3(\vex-\veu).\label{A.15}\ee

The factor $\bar Z$ and $\bar y_{123}$ in Eq.(\ref{45z}) contain
effective energies $\omega$ and $\Omega$. We now explain, how
these can  be computed, solving spin-independent Hamiltonian.

We start with the $Q\bar q$ system in its c.m. system. One has
(neglecting spin-dependent forces) \be H\psi\equiv
(\sqrt{\veq^2+m^2_q}+\sqrt{\veq^2+m^2_Q}+\Delta+V(r_{q\bar
Q}))\psi=M_{Q\bar q}\psi. \label{A.18}\ee Here $\Delta\equiv
\Delta_{SE}+\Delta_{string},~~ V(r_q\bar Q)=\sigma r_{q\bar
Q}+V_{coul}$ and $\omega\equiv \lan \sqrt{\veq^2+m_Q^2}\ran,$

$\Omega\equiv \lan \sqrt{\veq^2+m_q^2}\ran.$ Another way is to use
the einbein representation \cite{65,66} :  \be
\left(\frac{\veq^2+m^2_q}{2\omega}+\frac{\omega}{2}+\frac{\veq^2+m^2_Q}{2\Omega}+\frac{\Omega}{2}+
\Delta+V(r_{q\bar Q})\right)\psi=  M_{Q\bar q} (\Omega,\omega)
\psi \label{A.19}\ee with the subsequent optimization of $M_{Q\bar
q}(\Omega,\omega)$ using equation $\frac{\partial M}{\partial
\Omega}=\frac{\partial M}{\partial\omega}=0$, which yields
$\Omega=\Omega_0,~~\omega=\omega_0$ and it was shown that
$\omega_0,\Omega_0$ are close to the previous definition in terms
of averages \cite{66}, see \cite{67} for a review of string
Hamiltonian technic.

The spin-dependent forces are treated also in the FCM, as shown in
\cite{67}.

 The resulting spectrum for $Q\bar Q$ systems was obtained in
 \cite{61}, while that for $Q\bar q$ systems in \cite{57}.
It is important, that  the method contains minimal number of
parameters: current (pole) quark   masses, $(m_c=1.4$ GeV,
$m_b=4.8$ GeV, $m_u=m_d\approx 0$, $m_s=0.2$ GeV) string tension
$\sigma=0.18$ GeV$^2$, and $\alpha_s(q)=\frac{4\pi}{b_0ln
\frac{q^2+M^2_B}{\Lambda^2_{QCD}}}$ with $M_B\cong 1$ GeV and
$\Lambda^{(5fl)}_{QCD}=0.22$ GeV
In some recent applications also the flattening of the  confining
potential due to light pair creation was taken into account
\cite{68}, but we shall neglect this effect here.

With the given input parameters the resulting spectrum and
$\Omega,\omega$ are given in Table 1 for charmonium,  Table 2 for
bottomonium and  Table 3 for some heavy light mesons. Please note,
that values of $\Omega,\omega$ depend on the system, where a given
quark (antiquark) enter.

\vspace{3mm}
\newpage
{\bf Table 2}
 \begin{center}

{Values of $\Omega$ and  $\lan\vep^2\ran$ for $1^{--}$
states$^{*)}$ of charmonium, computed in FCM in \cite{18,68}
together with masses
  vs experimental data }

\vspace{3mm}

\begin{tabular}{|l|l|l|l|l|l|} \hline
 State& $J/\psi$&$\psi(2S)$& $\psi(3770)$&$\psi(3S)$&$\psi(4S)$\\
\hline $\Omega_R,$ GeV& 1.58& 1.647&1640&1.711&1.770\\\hline
$\lan \vep^2\ran_R$ GeV$^2$&0.569& 0.820&0.746&1.064&1.146\\
&&&&&1.300\\\hline

mass, MeV &3.090&3.675&3.800&4.094&4.442\\
theory $ M_{cog}$&&&&&\\\hline

mass, MeV&3067.8$\pm0.3$&3674.1$\pm1.0$&$3771.1\pm 2.4$&$4039\pm
1$&
$4421\pm 4 $ PDG\\
exper $M_{cog}$&&&$\theta=105^0$&&4411 (Belle)\\&&&&&4361?(
(Belle)\\
\hline $\lan p^2\ran_{EA}$ GeV$^2$&0.522&0730&0.731&0.940&1.141\\
\hline $\Omega_{EA}$, GeV& 1.60&1.65&1.65&1.70&1.75\\\hline

\end{tabular}

\vspace{3mm}

\end{center}

$^{*)} \Omega_R,\lan \vep^2\ran_R$ are computed via Salpeter
equation, Eq. (\ref{A.18}) while $\Omega_{EA}, \lan
\vep^2\ran_{EA}$ are computed in the einbein approximation,
Eq.(\ref{A.19}).

\vspace{3mm}
{\bf Table 3}
 \begin{center}
The same as in Table 1, but for   $1^{--}$ states of bottomonium,
as calculated in \cite{18,68}. $m_b=4.823,~~M_B=0.95$~~
$\Lambda_5(\overline{MS})=0.220;$ (all in GeV)  $\sigma=0.178$
GeV$^2$

\vspace{3mm}

\begin{tabular}{|l|l|l|l|l|l|} \hline
 State& $\Upsilon(1S)$&$\Upsilon(2S)$&$\Upsilon(3S)$&$\Upsilon(4S)$&$\Upsilon(5S)$\\
\hline $\Omega$, GeV& 5.021&5.026&5.056&5.088&5.120\\
\hline
$\lan \vep^2\ran$ GeV$^2$ EA&1.954&2.026&2.334&2.674&3.012\\
\hline

mass, GeV &9.460&10.010&10.356&10.633&10.873\\
theory {c.o.g.}&&&&&\\\hline $n^3S_1$ mass,
GeV&9.460&10.023&10.355(1)&10.579(1)&10.865(8)\\ \hline

\end{tabular}

\vspace{3mm}

\end{center}

\vspace{3mm}
\newpage
{\bf Table 4}
 \begin{center}

{Values of $\omega,\Omega$ and  $\lan\vep^2\ran$ for heavy-light
mesons
  (masses computed in \cite{57}  also given
  vs experimental data from PDG)}

\vspace{3mm}

\begin{tabular}{|l|l|l|l|l|} \hline
 Meson&$D$&$D_s$&$B$&$B_s$\\

\hline $\omega$, MeV&507&559&587&639\\
$\Omega$, MeV&1509&1515&4827&4830\\
$\lan \vep^2\ran,$ GeV$^2$& 0.273&0.290&0.359&0.383\\\hline

Mass, MeV&1869&1967&5279&5362\\
(theory)&&&&\\
exper., MeV&1869.3&1968.2&5279.0&5367.7\\
&$\pm0.4$& $\pm 0.4$& $\pm).5$&$\pm 1.8$\\
\hline

\end{tabular}

\vspace{3mm}

\end{center}


\vspace{2cm}

{\bf Appendix 2}\\

{\bf Normalization of  decay amplitudes in the relativistic representation }\\

 \setcounter{equation}{0} \def\theequation{A2.\arabic{equation}}

We start with the amplitude of two-body decay where initial and
final states are created by the current operators: initial  state
by $j_1=\bar \psi\Gamma_1\psi$ and final states
$j_i=\bar\psi\Gamma_i\psi,~~i=2,3$, while the pair creating
Lagrangian $\L=M_{br} (\bar \psi\Gamma_x\psi)$ acts at the point
$x$.

The Green's function for this amplitude in  the space-time can be
written as \be G_{123x}\equiv\lan 0|\prod_{i=1}^3j_i|0\ran= tr
(\Gamma_1S(1,2)\Gamma_2S(2,x)\Gamma_x S(x,3)\Gamma_3
S(3,1)).\label{a1'.1}\ee Here trace is over color, flavor and
Dirac indices, and  $S(i,k)$ is the quark propagator from the
point $i$ to the point $k$, see Fig.3. It is easy to see that
dimension of $\int d^4 x G_{123x}$ is $m^9$. To write propagators
in convenient relativistic form, we shall use the fact, that
between all quark lines in the diagram in Fig.3, corresponding to
the amplitude (\ref{a1'.1}), acts confinement, and therefore all
propagators are at the boundary of the film with string tension
$\sigma$, which creates effective mass $\bar \omega_q$ for all
quarks, including massless ones. This fact was established in
\cite{69}, where this mass was shown to be what is called the
constituent mass, and $\bar\omega_q$ was computed repeatedly and
accuracy was checked in \cite{66}.

The quark Green's function in the Fock-Feynman-Schwinger
representation has the form (in the Euclidean space-time) \be
S(x,y)=(m-\hat D)_x G(x,y),\label{a1'.2}\ee

\be G(x,y)=\int^\infty_0ds (D^4z)_{xy} e^{-K}\Phi_\sigma(x,y),
\label{a1'.3}\ee as it was shown in \cite{57}, one can rewrite
$G(x,y)$ identically as follows \be G(x,y) = \int
\frac{D\omega(D^3z)_{xy}}{2\bar \omega} e^{-K}\Phi_\sigma
(x,y)\label{a1'.5}\ee and the functional integral $(D\omega)$ has
the meaning of the averaging procedure over all possible values of
$\omega= \frac12\frac{dz_4}{d\tau}$, where $z_4(\tau)$ is  the
(Euclidean) time on the quark path at the proper time $\tau$. Note
that kinetic energy is $$K=\frac14 \int^s_0
\left(\frac{dz_\mu}{d\tau}\right)^2 d\tau+m^2s= \int^T_0 \left[
\frac{\omega}{2} \left(\left(\frac{dz_i}{dz_4}\right)^2+1\right)
+\frac{m^2}{2\omega}\right]dt$$ The averaging $D\omega$ with the
weight $\exp (-K)\Phi_\sigma$   finally yields $\bar \omega$ --
constituent mass.

Therefore $S(x,y)$ can be written as \be S(x,y) = \lan \bar Z
g(x,y)\ran_\omega,\label{a1'.5a}\ee where $\bar Z= \frac{(m-\hat
D)}{2\bar \omega}$, and \be g(x,y)= \int(D^3
z)_{xy}e^{-K}\Phi_\sigma(x,y).\label{a1'.6}\ee Consider now the
$q\bar q$ white system.\, e.g. the current-current correlator
\be\lan j(x_1) j_2(x_2)\ran =\lan tr \Gamma_1 S(x_1, x_2) \Gamma_2
S(x_2, x_1)\ran.\label{a1'.7}\ee

One can write, (neglecting spin splittings and hence second
exponent in (\ref{a1'.3})) \be \lan j_1j_2\ran =\lan \bar Y \int
(D^3z)_{x_1x_2} (D^3\bar z)_{x_1,x_2} e^{-K-\bar K}\lan W (x_1,
x_2)\ran_A\ran_{\omega_i}.\label{a1'.8}\ee Here
$$
\bar Y =\frac{tr \Gamma_1 (m_1-\hat D_1) \Gamma_2 (m_2-\hat
D_2)}{2\bar \omega_12\bar \omega_2}=$$ \be
=\frac{1}{4\bar\omega_1\bar\omega_{2}}tr (\Gamma_1(m_1-i\hat
p_1)\Gamma_2(m_2+i\hat p_2)).\label{a1'.9}\ee and $\lan W(x_1,
x_2)\ran$ is the Wilson loop average, with the closed loop along
trajectories of quark and antiquark.

In (\ref{a1'.8}) one can introduce relative and c.m. coordinates
$\veta,\verho$, and finally write
$$\int(D^3z)_{x_1x_2}(D^3\bar z)_{x_1x_2} e^{-K-\bar K}\lan
W\ran_A = \int (D^3\eta)_{0,0} (D^3\rho)_{x_1,x_2} e^{-K-\bar
K}\lan W\ran_A=$$
$$=\int \frac{d^3\veP}{(2\pi)^3}e^{i\veP(\vex_1-\vex_2)} \lan
\mathbf{0}|e^{-H(x_{24}-x_{14})}|\mathbf{0}\ran=$$ \be =\int
\frac{d^3\veP}{(2\pi)^3}e^{i\veP(\vex_1-\vex_2)} \sum_n
|\psi_n(0)|^2e^{-E_n\Delta T}.\label{a1'.10}\ee

Our important task is to go over from the point-point correlator
like $\lan j_1j_2\ran, $ to the hadron-hadron amplitude, which can
be done introducing the so-called decay constant $f^n_\Gamma$. For
the standard normalization of the hadron state (one hadron in the
volume $V_3=1$) one has\footnote{Note, that normalization of state
$|n\ran$ is (\ref{a1'.11}) corresponds to (\ref{a1'.10}) and
differs from \cite{57}.}\be \lan 0|j_\gamma|n,~~ \veP=0\ran
=\frac{\varepsilon_\gamma M_n
f^n_\Gamma}{\sqrt{2M_n}},\label{a1'.11}\ee with
$\varepsilon_\Gamma=1$ for  scalars and
$\varepsilon_\Gamma=\varepsilon_\mu^{(k)}$ for vectors.

Hence one can rewrite (\ref{a1'.8}) using (\ref{a1'.10}),
(\ref{a1'.11}) as \be\lan j_1j_2\ran= \bar Y \sum_n \int
\frac{d^3\veP}{(2\pi)^3}e^{i\veP(\vex_1-\vex_2)-E_n\Delta T}
\varepsilon_1\varepsilon_2\frac{E_n(f^n_\Gamma)^2}{2}.\label{a1'.12}\ee
Comparing (\ref{a1'.10}) and (\ref{a1'.12}) one obtains
$f^n_\Gamma$, (see \cite{57} for details and numerical estimates)
\be (f^n_\Gamma)^2=\frac{2N_cZ|\psi_n(0)|^2}{M_n}\label{a1'.13}\ee

In a similar way Eq.(\ref{a1'.11}) will help us now to define
hadron-hadron amplitudes by amputating the ``current-at-a-point''
matrix elements.

To this end we represent $G_{123x}$
 as follows (we omit integral
signs for brevity) \be G_{123x}=\bar Z_{123x} (D^3 z)_{12}
(D^3z)_{2x}(D^3z)_{x3}(D^3z)_{31} \prod D\omega_{ik}
e^{-K_{ik}}\lan W\ran\label{a1'.14}\ee

Here the subscript $(ik)$ runs over $(12),(2x),(x3)$ and (31), and
we shall omit the path integrals over $D\omega_{ik}$  and replace
it by the sign of averaging over $\omega$ of the whole expression,
$G$, e.g. $\lan G\ran_{\omega}$, which leads to the   appearing of
average values $\bar \omega$ in $\bar Z$, and the (multiparticle
in general) Hamiltonian $H$.

To introduce this Hamiltonian and corresponding eigenfunctions,
one should define the hyperplane, and we choose it as  a
 hyperplane at the time point $x_4$, which crosses path
(12) at space point $\veu$ and path (13) at the point $\vev$. One
can divide the paths at these points: \be (D^4z)_{12}=(D^3z)_{1u}
d^3\veu (D^3z)_{u2}, (D^3z)_{13} = (D^3z)_{1v}
d^3\vev(D^3z)_{03}\label{a1'.15}\ee and introduce c.m. and
relative coordinates in three regions (on three pieces of the
Wilson plane), denoted in Fig.3 by letters a,b, and c. For the
c.m. integral $(D^3\rho)_{\vey,\vez}$ one can use as in
(\ref{a1'.10}) the representation $\int \frac{
d^3\veP}{(2\pi)^3}\exp (i\veP(\vey-\vez))$.

Denoting the c.m. and relative coordinates in regions a,b,c as
$X_1, X_2, X_3$ and $\eta_1,\eta_2,\eta_3$ respectively, one can
rewrite $G_{123x}$ as $$ G_{123x} =\{\bar Z_{123x} \prod_{i=1}
D^3\eta_{ik}\frac{d^3P_i}{(2\pi)^3}d^3ud^3ve^{i\veP_1(\veX_1-\veX_{uv})}\times$$
\be\times e^{-i\veP_2(\veX_2-\veX_{ux}) -
i\veP_2(\veX_3-\veX_{xv})}e^{-K_{ik}} \lan
W\ran\}_{\omega_{ik}}\label{a1'.16}\ee

At this point  one can use as in (\ref{a1'.10}) the connection of
the einbein Hamiltonian with the path integral (see \cite{43} and
\cite{67} for details) \be\int
(D^3\eta_{ik})_{\ver_1,\ver_2}e^{-K_{ik}}\lan
W_{ik}\ran=\lan\ver_1|e^{-H_{ik}T}|\ver_2\ran.\label{a1'.17}\ee
Here $K_{ik}$ and $H_{ik}$ defined for a pair of quark paths $i$
and $k$, are \be K_{ik} = \int^T_{0_{dt}}\left\{
\frac{\omega_i+\omega_k}{2} +\frac{m^2_i}{2\omega_i}
+\frac{m^2_k}{2\omega_k}
+\frac{\omega_{ik}}{2}\left(\frac{d\eta_{ik}}{dt}\right)^2\right\}\label{a1'.18}\ee

\be
H_{ik}=\frac{w_i+\omega_k}{2}+\frac{m^2_i}{2\omega_i}+\frac{m^2_k}{2\omega_k}
+\frac{\vep^2_r}{2\omega_{ik}}+V(\eta_{ik},\veL^2).\label{a1'.19}\ee

Here $\omega_{ik}=\frac{\omega_i\omega_k}{\omega_i+\omega_k}$,and
the string potential $V$ was derived in \cite{69} and given in
Appendix D of \cite{57}.

The r.h.s. of (\ref{a1'.17})
 can be written as a spectral sum
 \be\lan
\ver_1 | e^{-H_{iK} T}| \ver_2\ran = \sum_n \psi_n(\ver_1)\psi^+_n
(\ver_2) e^{-E_nT}.\label{a1'.20}\ee

Using (\ref{a1'.16}) and keeping only fixed states $n_1, n_2, n_3$
in regions $a,b,c$ respectively, one gets for the Fourier
transform of $G_{123x}$

$$(\int G_{123x} d^4 x)_{\veP_1, \veP_2,\veP_3}= \bar Z_{123x} \int
d^4xd^3u d^3 v \Psi_{n_1}(0)\Psi_{n_1}^*(\veu-\vev
)\psi_{n_2}(\veu-\vex)\Psi^*_{n_2}(0)\times$$
$$
\times \psi_{n_3}(\vex-\vev)\psi^*_{n_3}(0)
e^{-i\veP_1\veX_{uv}+i\veP_2\veX_{ux}+i\veP_3\veX_{xv}}\times $$
\be \exp
\{-E_{n_1}(t_1-t_x)-E_{n_2}(t_x-t_2)-E_{n_3}(t_x-t_3)\}.\label{a1'.21}\ee
With the definition of c.m. coordinates $\veR,\veX_{uv}$ etc. as
in (\ref{A.14}), one can write \be d^4xd^3ud^3v=d^3\veR
d^3(\vev-\veu) d^3(\vex-\veu) dt_x\label{a1'.22}\ee integrating
over $d^3\ver$ one gets
$(2\pi)^3\delta^{(3)}(\veP_1-\veP_2-\veP_3)$, and integrating over
$dt_x$ yields the factor $2\pi\delta(E_1-E_2-E_3)$.

Now we
 are in position to amputate the ``current-at-a-point''  pieces
 $\psi_{n_1}(0)$ and go over from point-to-point correlators
 (Green's functions) to the hadron-hadron amplitudes. To this end
 one must replace every current vertex $\lan
 0|j_i\Lambda...|0\ran$ by $\lan
 0|j_i|n_i\veP_i\ran\lan n_i\veP_i|\Lambda...|0\ran$
and delete
 the factor on the left. Moreover, one deletes the energy factors
 $\exp(-\sum_i E_{n_i}t_i)$. As a result one obtains hadron-hadron
 amplitude

 $$ \lan n_1\veP_1|G|n_2\veP_2, n_3\veP_3\ran=$$
$$  =\frac{(\int G_{123x} d^4x)_{\veP_1,\veP_2,\veP_3}}{\prod_i\lan
 0|j_i|n_i,\veP_i\ran\exp(-\sum_i E_{n_i}t_i)}=$$
 \be
=\frac{(2\pi)^4}{\sqrt{N_c}}\delta^{(4)}(\mathcal{P}_1-\mathcal{P}_2-\mathcal{P}_3)J_{n_1,n_2,n_3}(\vep)_{n_3}
\label{a1'.23}\ee where we have defined \be \bar
y_{123}=\frac{\bar Z_{123x}}{\sqrt{\prod_{i=1,2,3}\bar
Z_i}},\label{a1'.24}\ee \be J_{n_1n_2n_3}(\vep)=\bar y_{123}\int
d^3(\vev-\veu)d^3(\vex-\veu)e^{i\vep\ver}\psi^*_{n_1}(\veu-\vev)\psi_{n_2}(\veu-\vex)\psi_{n_3}(\vex-\vev).
\label{a1'.25}\ee

Here $\vep=\frac12(\veP_1-\veP_2),~~\ver=c(\veu-\vev)$, and $c$ is
defined in (\ref{A.14}). The values of $\bar Z_i$ for different
$\Gamma_i=S.V,A,P$ can be obtained  as $\bar
Z_i=\frac{Y_{\Gamma_i}}{\omega_{1i}\omega_{2i}}$, and
$Y_{\Gamma_i}$ were calculated in \cite{57}.

Finally, the decay probability can be written as \be
dw=\frac{(2\pi)^4}{N_c}\delta^{(4)}(
\mathcal{P}_1-\mathcal{P}_2-\mathcal{P}_3)|J_{n_1n_2n_3}|^2\frac{d^3{\mathcal{P}}_2}{(2\pi)^3}
\frac{d^3{\mathcal{P}}_3}{(2\pi)^3}.\label{a1'.26}\ee

One can check, that $dw$ and $\Gamma=\int dw$ have correct
dimension, if taking into account that $ dim(\bar
y_{123})=dim\Gamma_x =[m],$ and $dim(J_{n_1n_2n_3})=[m^{-3/2}]$.

We note in conclusion, that one could define another normalization
of the bound states $|n,\veP>$, such that
$$AB=\sum_nA|n,\veP><n,\veP|B=\sum_n\frac{d^3\veP}{2E_n}A|n><n|B.$$
In this case the new amplitude
$$\lan n_1\tilde P_1|G|n_2\tilde P_2, n_3\tilde P_3\ran
=\sqrt{2E_{n_1}2E_{n_2}2E_{n_3}}\lan n_1 P_1|G|n_2P_2,
n_3P_3\ran$$ and in the definition of probability one should
divide (\ref{a1'.26}) by the factor $\prod^3_{i=1}(2E_{n_i})$,
which amounts to same $dw$ and $\Gamma$.

It is easy to see, that the emission of any number of additional
pions leads to an additional dimensionless factor
$\prod_i\frac{\exp i\veK_i\vex}{f_\pi\sqrt{2\omega_\pi(i)V_3}}$ in
the amplitude $J_{n_1n_2n_3}$ and appearance of additional
dimensionless factor $\prod_i\frac{V_3d^3\vek_i}{(2\pi)^3}$ in
$dw,$ therefore all normalization stays intact.

Finally we can compare $dw$ in (\ref{a1'.26}) with our expression
(47) for $\Gamma_n$. Identifying $J_{n_1n_2n_3}$ with
$(\sqrt{\gamma}J_{nn_2,n_3}(\vep))$ in (\ref{43}), one can see
that normalization of both expressions coincides. one can also
check, that in the nonrelativistic limit $\bar
y_{123}=\sqrt{\gamma}\bar Z_{nk}$, however for decay of heavy
quarkonia into 2 heavy-light mesons, $\bar y_{123}$ is twice as
big in the static limit due to $\sqrt{\bar Z_2\bar Z_3}$ in
(\ref{a1'.24}).


\vspace{2cm}

{\bf Appendix 3}\\

{\bf Kinematics of the dipion decays }\\

 \setcounter{equation}{0} \def\theequation{A3.\arabic{equation}}

We start with the two particle intermediate state, eg. $B\bar B$
or $B\bar B^*$, and define momenta of each hadron 1 or 2 as
$\veP_1, \veP_2$ and the total c.m. momentum is
$\veP=\veP_1+\veP_2$. Then one can write $\veP_1=\veP_Q+\vep_{\bar
q},~~\veP_2=\veP_{\bar Q}+\vep_{ q}$,
 and introduce relative momenta in hadrons 1 and 2:
 $\veq_1=\frac12(\veP_Q-\vep_{\bar q})$,~~$\veq_2=\frac12(\veP_{\bar Q}-\vep_{q})$.

 In the total c.m. system one has $\veP_1=-\veP_2\equiv \vep$, and
 hence $\vep_{\bar q}=\frac12 \vep-\bar q_1$, $\vep_{q}=-\frac12
 \vep-\veq_2$. Therefore the combination entering $\bar Z$ in
 (\ref{5.5}), $\vep_q-\vep_{\bar q}=-\vep+\veq_1-\veq_2$.
This is used in (\ref{5.9}) and one also obtains in (\ref{5.8})
that $\veq_1-\veq_2=0$.

Now consider the hypersurface at time $t_x$, when the  $q\bar q$
pair is created.  One has quarks $Q$ at coordinate $\veu, q\bar q$
at $\vex$ and $\bar Q$ at $\vev$. The $ \pi\pi$ system is also
created at $\vex$ with total momentum $\veK=\vek_1+\vek_2$, and
the total c.m. momentum $\veP=\veP_2+\veK$ and we keep notation
$\vep$ for  the relative  momentum, in the $B\bar B$ system, $
\vep=\frac12 (\veP_1-\veP_2)$. We define the c.m. coordinates of
hadrons $\veR_1,\veR_2$ and the total c.m.coordinate $\veR$, and
relative coordinates $\ver=\veR_1-\veR_2, \verho =\vex-\veR$,
where \be
\veR_1=\frac{\omega_1\vex+\Omega_1\veu}{\omega_1+\Omega_1},~
~\veR_2=\frac{\omega_2\vex+\Omega_2\vev}{\omega_2+\Omega_2},~~
\veR=\frac{E_1\veR_1+E_2\veR_2+(\omega_\pi(1)+\omega_\pi(2))\vex}{E_1+E_2+\omega_\pi(1)+\omega_\pi(2)}.\label{a2.1}\ee
Here $\omega_1=\omega_{\bar q}, \omega_2=\omega_q$ and similarly
for $\Omega_1,\Omega_2$. Writing the exponent of plane wave free
motion of hadrons 1,2 and two pions in terms of c.m. and relative
momenta $\vep$, and $\veR$, \be
\veP_1\veR_1+\veP_2\veR_2+\veK\vex= \veP\veR+\vep\ver+\vek\verho
\label{a2.2}\ee one arrives at the expressions \be
\vep=\frac{\alpha_2\veP_1-\alpha_1\veP_2}{\alpha_1+\alpha_2},~~\vek=\frac{1}{1-\alpha_3}
\veK,~~\alpha_i=\frac{\omega_i+\Omega_i}{E},i=1,2\label{a2.3}\ee
and $\alpha_3=1-\alpha_1-\alpha_2$, where $E$ is the total energy,
$ E=E_1+E_2+\omega_\pi(1)+\omega_\pi(2)$. In the equal mass (and
energy) case one arrives at the same equations  as before,
$\alpha_1=\alpha_2, \vep=\frac12(\veP_1-\veP_2)$ and
$\vep_q-\vep_{\bar q}=\veq_1-\veq_2-\vep.$

However now integrating as in (\ref{44z}) over the coordinate
$\vex$ where the $q\bar q$ and $\pi\pi$ pairs are emitted one has
instead of previous case without pion emission,
$\veq_1-\veq_2=\veK$(we neglect small corrections of the order of
$\frac{\omega_\pi}{m_Q}$).

Hence for the diagram with the 2$\pi$ emission one has $\bar Z\sim
(p_q-p_{\bar q})_i=K_i-p_i$ as used in (\ref{6.11}).

In the second part of the diagram (the r.h.s. part of Fig.) pions
are not emitted at the point $\vey$, and $\veq'_1-\veq'_2=0$,
hence there $\bar Z$ is the same as in the pionless decay case.

We turn now to the 3 body phase space of $X(n')\pi\pi$ and useful
coordinates on the Dalitz plot. We choose as such the standard
quantities:  invariant  dipion mass $M^2_{\pi\pi} \equiv q^2 =
(k_1+k_2)^2=(\omega_1+\omega_2)^2-\veK^2$, and $\cos \theta$,
where $\theta$ is the angle of $\pi^+$ in the c.m. of $\pi^+\pi^-$
with  respect to the direction of incident quarkonium. One has \be
\cos\theta=
\frac{q}{\sqrt{q^2-4m^2_\pi}}\frac{M^{'2}-M^2-q^2+4M\omega_2}{\sqrt{[M^2-(q+M')^2][M^2-(M'-q)^2]}}.\label{A2.4}\ee

The pion energies $\omega_1 $ and $\omega_2$  can be written as
\be \omega_{1,2} =\tilde c\mp \tilde
d\cos\theta=\frac{(M+M')\Delta M+q^2}{4M}\mp \frac{(M+M')}{4M}
\frac{\sqrt{q^2-4m^2_\pi}{\sqrt{(\Delta
M)^2-q^2}}}{q}\cos\theta.\label{A2.5}\ee

Therefore the sum $\omega_1+\omega_2$ and $\veK^2$ are \be
\omega_1+\omega_2=\frac{q^2+M^2-M^{'2}}{2M},~~
\veK^2=\frac{((\Delta
M)^2-q^2)((M+M')^2-q^2)}{4M^2}\label{A2.6}\ee and the combination
appearing in (\ref{6.12}), is \be \vek^2_1+\vek^2_2= 2(\tilde
c^2+\tilde d^2\cos \theta)-2m^2_\pi\equiv \alpha+\gamma
\cos^2\theta\label{A2.7}\ee where we have defined for the reaction
$X(n)\to X(n')\pi\pi; M=$ mass of $X(n)$, $M'=$  mass of $X(n'),
~~ \Delta M= M-M'$.
\newpage

\vspace{2cm}

{\bf Appendix 4}\\

{\bf The overlap integrals $J^{(k)}_{nn'}$ and $p^{(k)}_{nn'}$ }\\

 \setcounter{equation}{0} \def\theequation{A4.\arabic{equation}}

One starts with the SHO wave functions, which can be written as
$\Psi_{n}(q,\beta) = \mathcal{P}_n(q)e^{-q^2/2\beta^2},  n
=1,2,3,..$ for $(nS)$  states \be \mathcal{P}_1(q)
=\left(\frac{2\sqrt{\pi}}{\beta}\right)^{3/2}
c_1,~~\mathcal{P}_2(q) =c_2
\left(\frac{2\sqrt{\pi}}{\beta}\right)^{3/2}\left(1-\frac23\left(\frac{q}{\beta}\right)^2\right)
,\label{A4.1}\ee
$$
\mathcal{P}_3(q) =\left(\frac{2\sqrt{\pi}}{\beta}\right)^{3/2} c_3
\left(\frac{15}{4} -5\left(\frac{q}{\beta}\right)^2 +
\left(\frac{q}{\beta}\right)^4\right); ~~c_1=1,c_2=\sqrt{\frac32},
c_3=\sqrt{\frac{2}{15}}.$$

The overlap integrals of the $Q\bar Q nS$ state and the $n=1$
state of heavy-light mesons $(B\bar B$ or $BB^*$ for $\Upsilon
(nS))$, are
$$ e^{-\frac{c^2\vep^2}{\Delta}} I_n(p) =\int \frac{d^3q}{(2\pi)^3} \Psi_n (\veq + c\vep,
\beta_1) \Psi^2_1 (q, \beta_2)=$$ \be =\left(
\frac{2\sqrt{\pi}}{\beta_1}\right)^{3/2} c_nc^2_1
e^{-\frac{c^2\vep^2}{\Delta}}\{\}_n \lambda^{3/2}, ~~
\lambda=\frac{2\beta^2_1}{2\beta^2_1+\beta^2_2},\label{A4.2}\ee
\be \{\}_1 =1 ; \{\}_2 =y-\frac83
\frac{c^2\vep^2\beta^2_1}{\Delta^2}; ~~
c=\frac{\Omega}{\Omega+\omega}\approx 1\label{A4.3}\ee
$$\{\}_3 =\frac{15}{4}y^2 - 5y \left(\frac{\lambda c
p}{\beta_1}\right)^2+\left(\frac{\lambda c p}{\beta_1}\right)^4.$$

 According to (\ref{93}) the overlap integrals $\mathcal{F}^{(k)}_{nn'}$
 with $n,n'=1,2,3,4$ for 1S,2S,3S,4S states respectively can be
 written as
 $$\mathcal{F}^{(k)}_{nn'}=\int\frac{d^3\vep}{(2\pi)^3}e^{-\frac{p^2}{\beta^2_0}}p^{2k}I_n
 (p)I_{n'}(p)=$$
 \be= c_nc_{n'}{(tt')^{3/2}
 \beta_0^{2k}}p^{(k)}_{nn'},~~
 c_1=1,~~c_2=\sqrt{\frac32},~~c_3=\sqrt{\frac{2}{15}}\label{a3.1}\ee.

 Using $I_n(p)$ defined in (\ref{6.19}), one obtains
$$
p_{21}^{(0)}=y-t^2,~~p_{31}^{(0)}=\frac{15}{4} (y-t^2)^2,$$ \be
p_{32}^{(0)}=-\frac54(-3y^2y'+6t^2yy'-3y't^4+3y^2t^{'2}-10yt^2t^{'2}+7t^{'2}t^4),\label{a3.2}\ee
$$
p_{21}^{(1)}=-\frac32(-y+\frac53
t^2),~~p_{31}^{(1)}=\frac{15}{8}(7t^2-3y)(t^2-y)$$
$$
p_{32}^{(1)}=-\frac{15}{8}
(-3y^2y'+10yy't^2+7t^4+5y^2t^{'2}+\frac{70}{3}t^2t^{'2}+21t^{'2}t^4)$$

Here we have defined in $p^{(k)}_{32}$: $\beta_1=\beta(3S);
\beta'_1=\beta(2S),~~\beta_2=\beta(B)\approx $  $\beta (B^*)$ or
$D,D^*$.

$$t=\frac{2\beta_1\beta_0}{\Delta},~~t'=\frac{2\beta'_1\beta_0}{\Delta'},~~
\Delta=2\beta^2_1+\beta^2_2,~~\Delta'=2\beta^{'2}_1+\beta^2_2$$

$$y=\frac{2\beta^2_1-\beta^2_2}{\Delta},~~y'=\frac{2\beta^{'2}_1-\beta^2_2}{\Delta'},~~
\beta^{-2}_0=\Delta^{-1}+\Delta^{'-1}$$.

For $p^{(k)}_{31}$ $\beta_1=\beta_1(3S),~~\beta'_1=\beta_1(1S)$
etc.

We are defining all $\beta_1(nS)$ through m.s.r. of the
corresponding $Q\bar Q$ states, using SHO wave functions,
$\beta_1(nS)=\sqrt{\frac{(4n-1)}{2\lan r^2\ran}}$,and taking $\lan
r^2\ran_{nS}$ for $\Upsilon$ from \cite{18,67} one has $\lan
r^2\ran_{nS}=(0.2$ fm)$^2,~(0.5$ fm)$^2$, (0.7 fm)$^2,~(0.9$
fm)$^2$, (1.1 fm)$^2$ for $n=1,2,3,4,5$ respectively, which gives
$\beta_1(nS)=1.22$ GeV, 0.75 GeV, 0.67 GeV,0.61 GeV, 0.56 GeV.

For $B,B^*(n=1)$ one has $\lan r^2\ran_{1S}=(0.5$ fm$)^2$,
\cite{57} and therefore $\beta_2=0.49$ GeV; for $D,D^*(\lan
r^2\ran_D=(0.58$ fm$)^2$ and $\beta_2=0.42$ GeV.

$\beta_0$ is defined by both $nS$ and $n'S$, therefore
$\beta_0(3S,1S)=0.92$ GeV, $\beta_0(2S,1S)=0.96$ GeV,
$\beta_0(3S,2S)=0.79$ GeV.

Now one can define the parameter
$\rho_{nn'}=\frac{p_{nn'}^{(1)}}{p^{(0)}_{nn'}}$ for
$n,n'=2,1;3,1;3,2$. Using (\ref{a3.2}) and values of $\beta_i$
quoted above, one has
$$
\rho_{21}=\frac32 \frac{(-y+\frac53 t^2)}{(-y+t^2)}=3.74 ;~~
\rho_{31} =\frac{(3y^2+7t^2)}{2(-y+t^2)}=5.33 $$ \be
\rho_{32}=6.79.\label{a3.3}\ee

 To estimate (\ref{6.22}) one extracts $\Delta M^*_{nn'}$ and
$\Delta M_{nn'}$ given in the Table 1 for the $\Upsilon (nS) \to
\Upsilon(n'S)\pi\pi$ transitions with $BB^*$ and $BB$ intermediate
sates respectively.

\newpage

\vspace{3mm}
{\bf Table 5}
 \begin{center}

{The values of $\Delta M^*_{nn'}$ and $\Delta M_{nn'}$(in GeV) for
lowest $nS$ states.}

\vspace{3mm}

\begin{tabular}{|l|l|l|l|} \hline

State&$2S$&$3S$&$4S$\\
\hline $\Delta M^*_{nn'}(E_{ns})$&0.58&0.25&0.025\\

\hline $\Delta M_{nn'}(E_{ns})$&0.54&0.205&-0.02\\

\hline

\end{tabular}

\end{center}

Note, that another nearest threshold for $\Delta M_{nn'}$ is
$B^*B^*$, which increases $\Delta M_{nn'}$ by 0.09 GeV. For a
rough estimate of $\lan \omega_1\ran, ~~ \lan
\omega_1+\omega_2\ran$ we can use the values of anailable phase
space $\Delta M= M_n-M_{n'}$ for $2S\to 1S, 3S\to 1S$ and $3S\to
2S$ cases respectively, and the values of $\lan \omega_1\ran$ are
given in the Table 6.

\vspace{3mm}
{\bf Table 6}
 \begin{center}
\vspace{3mm}

\begin{tabular}{|l|l|l|l|} \hline

decay&$2S\to 1S$&$3S\to 1S$&$3S\to 2S$\\
\hline $\Delta M$,GeV&0.56&0.895&0.332\\
$\lan \omega_i\ran$, GeV&0.28&0.44&0.166\\
$\beta^2_0$& 0.96&0.84&0.626\\
$\alpha$&0.12&0.36&0.015\\
$\gamma$& 0.16&0.4&0.055\\
$(c_nc_{n'} p^{(0)}_{nn})^2$ &0.39&0.257&0.17\\
$\rho_{nn'}$ & 3.74&5.33& 6.79\\
\hline

\end{tabular}

\end{center}

\vspace{3mm}
{\bf Table 7}\\
 Parameters of the $\Upsilon(nS)$ SHO eingefunctions and overlap integrals fitted
 to the known r.m.s.radii of $nS$ states.
\vspace{3mm}
\begin{center}
\begin{tabular}{|l|l|l|l|} \hline

$nS$&$ 1S$&$2S$&$3S$\\
\hline $\beta_1$&1.22&0.75&0.67\\
\hline $\Delta$ & 3.22&1.375&1.15\\
\hline $y$ &0.384&0.636&0.56\\
\hline

\end{tabular}

\end{center}

\vspace{3mm}
{\bf Table 8}\\
The same as in Table 7 for intermediate quantities and the final
one $\eta$ in two lines; $\eta_{SHO}$ for direct calculation, Eq.
(108) and $\eta_{SHO}^{AZI}$ the $AZI$ -improved values, as
discussed in the text.
 \begin{center}
\vspace{3mm}

\begin{tabular}{|l|l|l|l|} \hline

$nS\to n'S$& 21&31&32\\
\hline
$t$& 1.07&1.08&0.92\\
\hline
$\beta^2_0$&0.96&0.84&0.626\\
\hline
$\left(tt'\right)^3$& 0.5&0.416&0.5\\
\hline

$\zeta^2$ & $0.2$&$0.107$ &$ 0.085$\\
\hline $\xi^* $& 0.638&0.657&0.813\\
\hline $\xi$& 0.824&0.969&1.033\\ \hline $\tilde \xi$&0.56&0.5&0.761\\
\hline $\eta_{SHO}$&$<0.45$&$<0.27$&-3.66\\
\hline$\eta^{AZI}_{SHO}$&0.051&0.39&-3.2\\
\hline $\eta_{\exp}(fit)$ &0&0.52&-2.7\\

\hline

\end{tabular}

\end{center}

\vspace{1cm}

{\bf Figure captions}

\begin{description}
    \item[]Fig.1.~Light quark pair creation inside heavy
    quarkonia.
    \item[]Fig.2.~Emission of two pions from the light quark
    loop: a) two-pion emission b) successive one-pion emission.
    \item[]Fig.3.~The  connected 4-point Green's function for
    the decay of heavy quarkonium into two heavy-light mesons.
    \item[]Fig.4.~Pictoral image of the successive pion
    emission with heavy-light mesons in the intermediate state.
    \item[]Fig.5.~The same as in Fig. 4, but with two-pion emission
    at one point.
    \item[]Fig.6.~Experimental data of the Cleo Collaboration
    from [42] with the theoretical parametrization as in Eq.
    (115), $ \frac{dw}{dq}=$ const (phase space) $|\eta-x|^2$,
    with $\eta(fit)$ given in Table 1 in comparison to theoretical
    predictions.

        \item[]Fig.7.~A diagram of two heavy-light meson
        contribution to the heavy quarkonium Green's function.
\end{description}

\end{document}